\documentclass[12pt]{article}
\usepackage{amsmath,amsthm,amssymb,amsfonts}
\usepackage{graphicx}
\usepackage{float}
\usepackage{enumerate}
\usepackage{natbib}
\usepackage{url}
\usepackage{bm}
\usepackage{mathrsfs}
\usepackage{multirow}
\usepackage{booktabs}
\usepackage{dsfont}
\usepackage{csquotes}
\usepackage{subcaption}
\usepackage{appendix}
\usepackage{placeins}
\usepackage[labelfont=bf]{caption}
\usepackage[colorlinks,citecolor=blue,urlcolor=blue,hyperfootnotes=false]{hyperref}

\newcommand{\blind}{0}

\graphicspath{{./frames/}}

\numberwithin{equation}{section}
\setcitestyle{open={(},close={)}}
\allowdisplaybreaks

\theoremstyle{definition}
\newtheorem{theorem}{Theorem}
\newtheorem{lemma}{Lemma}

\def\spacingset#1{\renewcommand{\baselinestretch}{#1}\small\normalsize} \spacingset{1}

\begin{document}
	
	\if0\blind
	{
		\title{\bf Asymmetric Huber Periodogram}
		\author{Tianbo Chen\thanks{School of Big Data and Statistics, Anhui University, Hefei, China.}\\
			Anhui University\\
			and \\
			Xiaojun Song\thanks{Corresponding author. Guanghua School of Management, Peking University, Beijing, China. Email: \texttt{sxj@gsm.pku.edu.cn}.} \\
			Peking University}
		\maketitle
	} \fi
	
	\if1\blind
	{
		\bigskip
		\bigskip
		\bigskip
		\begin{center}
			{\LARGE\bf Asymmetric Huber Periodogram}
		\end{center}
		\medskip
	} \fi
	
	\bigskip
	\begin{abstract}
		This paper introduces a novel spectral M-estimator, called the asymmetric Huber periodogram (AHP), as a generalization of the ordinary periodogram (PG), the quantile periodogram (QP), and the Huber periodogram (HP). The AHP is constructed via trigonometric asymmetric Huber regression (AHR), in which a specially designed loss function replaces the squared $\ell_2$ loss used to define the PG. Relative to the QP, the AHP can be more computationally efficient, and relative to the HP and PG, it provides a more comprehensive characterization by examining the data across the range of the asymmetry parameter. We establish the theoretical properties of the AHP and investigate its relationship with the corresponding asymmetric Huber spectrum (AHS). Building on the asymptotic theory, we develop confidence intervals (CIs) for the AHS and propose a Fisher-type test. Simulation studies and two applications %involving financial and physiological data
		further demonstrate the AHP's effectiveness in detecting hidden periodicities, robustness to outliers, and utility for time series clustering.
	\end{abstract}
	
	\vfill
	
	\newpage
	\spacingset{1.8} % DON'T change the spacing!
	\section{Introduction}
	The PG is a nonparametric estimator of the power spectrum that plays a crucial role in spectral analysis and is widely applied across various fields \citep{caiado2006periodogram,polat2007classification,baud2018multi,euan2018hierarchical,chen2021clustering}. The PG is constructed via trigonometric ordinary least squares (OLS) regression. However, OLS has well-known limitations in terms of robustness and efficiency, particularly when errors are asymmetric or heavy-tailed. Moreover, because OLS targets the conditional mean, it cannot characterize distributional features of the response distribution, %which in turn 
	adversely affecting the performance of the PG \citep{bloomfield2004fourier}.
	Therefore, alternative regression techniques have been developed to overcome these issues.
	
	The pioneering work of \cite{koenker1978regression} introduced quantile regression (QR). By replacing the $\ell_{2}$ loss with an asymmetric $\ell_{1}$ loss, QR provides a more complete picture of the relationship between the response variable and the covariates across different quantile levels, while also demonstrating strong robustness against outliers and heavy-tailed distributions.   However, QR is not free of limitations: the non-smooth check (loss) function imposes both theoretical and computational burdens and suffers from efficiency losses under Gaussian or homoscedastic errors.   To alleviate these issues, asymmetric least squares (ALS) regression, also known as expectile regression (ER), was proposed by \cite{newey1987asymmetric}. Its quadratic loss mitigates some of the difficulties of QR but at the cost of reduced robustness.  A comprehensive comparative analysis of quantiles and expectiles is provided in \cite{efron1991regression,jones1994expectiles,yao1996asymmetric,waltrup2015expectile}. 
	QR and ER, as well as their extensions, have been widely used across various scientific fields \citep{koenker1999goodness,koenker2005quantile,machado2005counterfactual,gu2016high,ziegel2016coherence,bellini2017risk,jiang2017expectile,guler2017mincer,daouia2018estimation,xu2020mixed,dawber2022expectile}. As a generalization of QR and ER, the $\ell_p$ quantile was introduced by \cite{chen1996conditional} and further studied by \cite{jiang2021k}.
	
	Building on these regression frameworks, researchers have developed the corresponding periodograms. One innovative development is the QP \citep{li2012quantile} constructed via trigonometric QR, which demonstrates its ability to detect hidden periodicities. In large samples, the QP is associated with the so-called quantile spectrum, a scaled version of the ordinary power spectrum of the level-crossing process. The Laplace periodogram (LP) \citep{li2008laplace} arises as the special case for the 0.5 quantile level. Related work on the QP includes \cite{lidet,li2014quantile,dette2015copulas,kley2016quantile,birr2017quantile,meziani2020penalised,chen2021semi,li2023quantile}. \cite{chen2026expectile} further proposed the EP and provided a comprehensive comparison with the QP. For more robust periodograms, \cite{wen2021robustperiod} proposed the HP, and \cite{thieler2016robper} developed the R package \texttt{RobPer}, which enables the combination of different robust regression techniques \citep{rousseeuw1984robust,oh2004period} to compute robust periodograms. 
	
	Beyond $\ell_p$ quantile regression, the asymmetric Huber regression (AHR) \citep{allende2006m} also strikes a balance between robustness and efficiency by using piecewise linear and quadratic losses, with a threshold parameter to control the ratio of the two losses. The AHR is closely related to M-estimation with a continuously differentiable objective function and, unlike convolution-smoothed quantile regression \citep{horowitz1998bootstrap}, it maintains exact robustness by virtue of the bounded derivative $\dot\rho_{\alpha,\psi}$ beyond the threshold. In addition to the original Huber loss \citep{huber1964robust}, the AHR uses an asymmetry parameter that, similar to quantile and expectile regression, enables a comprehensive analysis of the data distribution. In this paper, we apply trigonometric AHR to define the AHP. We show that the AHP retains key properties of the PG, LP, and HP as a frequency-domain representation of serial dependence, while offering broader distributional insight by varying the asymmetry and threshold parameters. Relative to the EP, truncation at the threshold parameter in the AHP improves robustness while maintaining efficiency at larger thresholds. Computationally, the AHR objective is solved via iteratively reweighted least squares (IRLS), which converges rapidly due to the quadratic behavior near the origin; this contrasts with the linear programming required for the QP \citep{koenker2005quantile}.

	The remainder of the paper is organized as follows. Section \ref{EP} introduces the AHP. Section \ref{AS} develops large-sample theory and derives the asymptotic distribution of the AHP, along with its main properties. Section \ref{sim} presents simulation studies comparing different periodograms. Section \ref{APP} presents two applications, highlighting the AHP's ability to detect hidden periodicities, its robustness to outliers, and its utility for time series clustering. Section \ref{CO} concludes the paper. %with limitations and directions for future work.
	
	\section{Asymmetric Huber Periodogram}\label{EP}
	
	\subsection{Ordinary Periodogram}
	Let $\omega_{\nu} = 2\pi \nu / n$ for $\nu = 0,1,\ldots,n-1$ denote the Fourier frequencies, and let $\{y_t:t=1,\ldots,n\}$ be a real-valued time series. The PG is defined as 
	\begin{equation}\label{pg1}
		I_{n}(\omega_{\nu}) = n^{-1} \left| z_{n}(\omega_{\nu}) \right|^{2} 
		= n^{-1} \left| \sum_{t=1}^{n} y_t \exp(- {\rm i} \omega_{\nu} t) \right|^{2},
	\end{equation}
	where $z_{n}(\omega_{\nu})$ denotes the discrete Fourier transform (DFT) of $\{y_t\}$ and ${\rm i}=\sqrt{-1}$.
	
	An equivalent representation follows from an OLS regression formulation. Let
	$
	\tilde{\bm{\beta}}_n(\omega_{\nu}) = [\tilde{\beta}_1(\omega_{\nu}),\tilde{\beta}_2(\omega_{\nu}),\tilde{\beta}_3(\omega_{\nu})]^{\top},
	$
	which minimizes
	\begin{equation}\label{ols}
		\tilde{\bm{\beta}}_n(\omega_{\nu}) := \arg\min_{\bm{\beta} \in \mathbb{R}^3} \sum_{t=1}^{n} \left(y_t - \mathbf{x}_{nt}^{\top} \bm{\beta}\right)^2,
	\end{equation}
	where the regressor $\mathbf{x}_{nt}$ is defined as
	\begin{equation}\label{xt}
		\mathbf{x}_{nt}=\mathbf{x}_{nt}(\omega_{\nu}): = \left[1, \cos(\omega_{\nu} t), \sin(\omega_{\nu} t)\right]^{\top}.
	\end{equation}
	Then, it can be verified that the PG in (\ref{pg1}) satisfies $I_{n}(\omega_{\nu})=n\{\tilde{\beta}_2(\omega_{\nu})^{2} + \tilde{\beta}_3(\omega_{\nu})^{2}\}/4.
	$
	In practice, one can set $I_n(0)=0$, which corresponds to computing the periodogram from the demeaned series $y_t - \bar{y}$, where $\bar{y} = n^{-1}\sum_{t=1}^n y_t$ denotes the sample mean.

	\subsection{Asymmetric Huber Periodogram}
\begin{figure}[ht]
		\centering
		
		\includegraphics[width=0.7\textwidth]{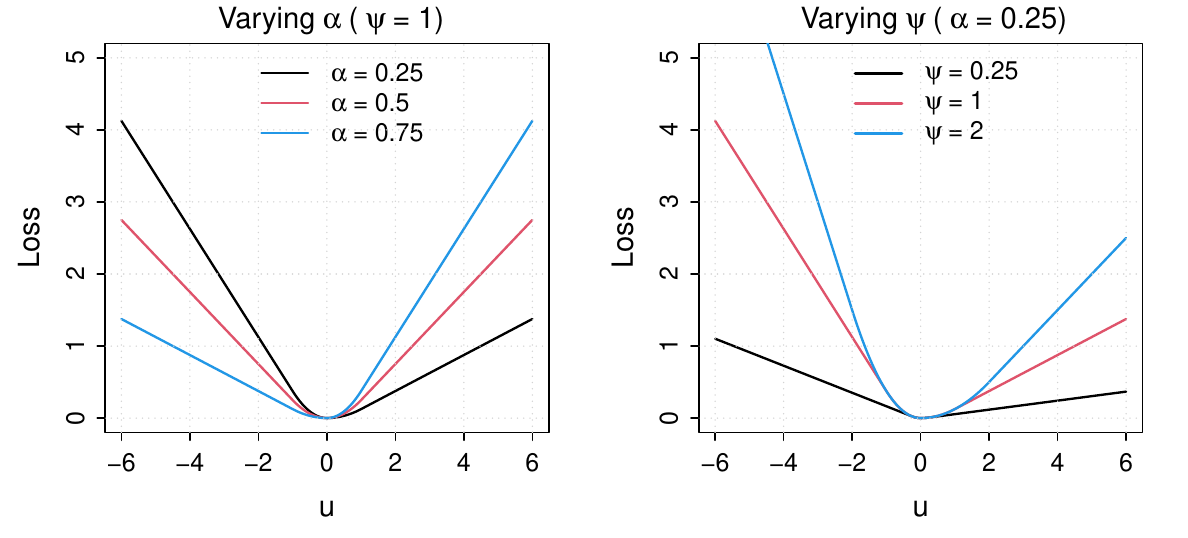}
		
		\caption{The AHL under different values of $\alpha$ and $\psi$.}
		\label{loss}
	\end{figure}

	For any fixed asymmetry parameter $\alpha \in (0,1)$ and threshold parameter $\psi > 0 $, define the asymmetric Huber loss (AHL)\footnote{Note that some literature, such as \cite{gupta2020robust} and \cite{hazarika2023robust}, uses a different version, in which the loss function is symmetric when $|u| \leq \psi$.} as
	\begin{equation}\label{ahl}
		\rho_{\alpha,\psi}(u) =
		\begin{cases}
			|\alpha - \mathbf{1}(u < 0)| \cdot \frac12 u^2, & |u| \le \psi, \\
			|\alpha - \mathbf{1}(u < 0)| \cdot \psi ( |u| - \frac12 \psi ), & |u| > \psi,
		\end{cases}
	\end{equation}
	where $\mathbf{1}(\cdot)$ denotes the indicator function. The $\alpha$-th asymmetric Huber quantile (AHQ) of $y_t$ is defined as
	\begin{equation}\label{mu}
		\mu(\alpha,\psi) := \arg\min_{\mu \in \mathbb{R}} \mathrm{E} \left\{ \rho_{\alpha,\psi}(y_t - \mu) \right\},
	\end{equation}
	with the normal equation
	\begin{equation}\label{ne}
		\mathrm{E} \left\{ \dot{\rho}_{\alpha,\psi}(y_t - \mu(\alpha,\psi)) \right\} = 0,
	\end{equation}
	where
	\begin{equation}\label{diff}
		\dot{\rho}_{\alpha,\psi}(u)
		=
		\left|\alpha-\mathbf{1}(u<0)\right|
		\begin{cases}
			u, & |u|\leq\psi,\\
			\psi\,\operatorname{sgn}(u), & |u|>\psi.
		\end{cases}
	\end{equation}
	Fig.~\ref{loss} illustrates the effect of the asymmetry parameter $\alpha$ and the threshold parameter $\psi$ on the shape of the AHL.

	Substituting the AHL (\ref{ahl}) into (\ref{ols}) yields
	\begin{equation}\label{er}
		\hat{\bm{\beta}}_n(\omega_{\nu},\alpha,\psi) := \arg\min_{\bm{\beta} \in \mathbb{R}^3} \sum_{t=1}^{n}\rho_{\alpha,\psi}\left( y_t - \mathbf{x}_{nt}^{\top} \bm{\beta}\right),
	\end{equation}
	where $\hat{\bm{\beta}}_n(\omega_{\nu},\alpha,\psi) := [ \hat{\beta}_1(\omega_{\nu},\alpha,\psi) , \hat{\beta}_2(\omega_{\nu},\alpha,\psi) , \hat{\beta}_3(\omega_{\nu},\alpha,\psi) ] ^ {\top}.$
	The raw AHP is defined as
	\begin{equation}\label{ahp}
		{\rm AHP}_{n}(\omega_{\nu},\alpha,\psi):=\frac{n}{4} \left\{ \hat{\beta}_2(\omega_{\nu},\alpha,\psi)^2 +\hat{\beta}_3(\omega_{\nu},\alpha,\psi )^2 \right\}, \quad \text{for }\nu=1,\ldots,\lfloor\frac{n-1}{2}\rfloor.
	\end{equation}
	When $\omega_{\nu}$ equals $0$  or $\pi$, we modify the regressors accordingly. For $\omega_{\nu} = \pi$, define  
	$
	\mathbf{x}_{nt}(\pi) := [ 1 , \cos(\pi t)] ^{\top}, 
	$
	and let the resulting AHR solution be  
	$
	\hat{\bm{\beta}}_n(\pi, \alpha,\psi) :=[ \hat{\beta}_1(\pi, \alpha,\psi) , \hat{\beta}_2(\pi, \alpha,\psi) ]^{\top}.
	$
	For $\omega_{\nu} = 0$, let $\hat{\beta}_1(0, \alpha,\psi)$ denote the AHR solution with the regressor $\mathbf{x}_{nt} := 1$. Then,
	\[
	{\rm AHP}_{n}(\omega_{\nu},\alpha,\psi) := \left\{
	\begin{array}{ll}
		n|\hat{\beta}_{1}(0,\alpha,\psi)|^{2}, & \omega_{\nu}=0, \\
		n|\hat{\beta}_{2}(\pi,\alpha,\psi)|^{2}, & \omega_{\nu}=\pi \text{ (if $n$ is even)}.
	\end{array}
	\right. 
	\]
	In applications focusing on serial dependence, we can set
	$
	\operatorname{AHP}_n(0,\alpha,\psi)=0,
	$
	analogously to computing the ordinary periodogram from a demeaned
	series.
	
	The AHP is a generalization of the following periodograms:
	\begin{itemize}
		\item	Huber periodogram, when $\alpha = 0.5$, 
		\item	Ordinary periodogram, when $\alpha = 0.5$ and $\psi \to \infty$, 
		\item   Expectile periodogram, when  $\psi \to \infty$,
		\item   After rescaling the loss function by $1/\psi$, the asymmetric Huber
		loss converges pointwise to the quantile check loss as
		$\psi \downarrow 0$. Accordingly, under suitable regularity conditions, the
		AHP reduces to the QP in the limit. Furthermore,
		when $\alpha=0.5$, the QR estimator becomes the median regression
		estimator, so that the limiting AHP reduces to the LP.
	\end{itemize}
	Table~\ref{tab:comparison} summarizes how each periodogram is recovered from the AHP by appropriate choices of $(\alpha,\psi)$.
	
	\begin{table}[ht]
		\centering
		\caption{Recovery of existing periodograms from the AHP.}
		\label{tab:comparison}
		\spacingset{1}
		\footnotesize
		\setlength{\tabcolsep}{4pt}
		\begin{tabular}{lccc}
			\toprule
			Periodogram & $\alpha$ & $\psi$ & Reference \\
			\midrule
			Ordinary periodogram (PG) & $0.5$ & $\infty$ & \cite{brockwell1991time} \\
			Laplace periodogram (LP) & $0.5$ & $0^+$ (limit) & \cite{li2008laplace} \\
			Huber periodogram (HP) & $0.5$ & $\psi\in(0,\infty)$ & \cite{wen2021robustperiod} \\
			Quantile periodogram (QP) & $\alpha\in(0,1)$ & $0^+$ (limit) & \cite{li2012quantile} \\
			Expectile periodogram (EP) & $\alpha\in(0,1)$ & $\infty$ & \cite{chen2026expectile} \\
			Asymmetric Huber periodogram (AHP) & $\alpha\in(0,1)$ & $\psi\in(0,\infty)$ & This paper \\
			\bottomrule
		\end{tabular}
		\spacingset{1.8}
	\end{table}
	
	The AHP can also be expressed in Fourier-transform form. Given the AHR solution $	\hat{\bm{\beta}}_n(\omega_{\nu},\alpha,\psi)$, the Asymmetric Huber discrete Fourier transform (AHDFT) of $\{y_{t}\}$ is defined as
	\[
	z_{n}(\omega_{\nu},\alpha,\psi):=\left\{
	\begin{array}{ll}
		n\hat{\beta}_{1}(0,\alpha,\psi), & \omega_{\nu}=0, \\
		n\hat{\beta}_{2}(\pi,\alpha,\psi), & \omega_{\nu}=\pi \text{ (if $n$ is even)}, \\
		(n/2)\{\hat{\beta}_{2}(\omega_{\nu},\alpha,\psi)-i\hat{\beta}_{3}(\omega_{\nu},\alpha,\psi)\}, & \omega_{\nu} \in (0, \pi).
	\end{array}
	\right. 
	\]
	Then, it can be verified that the AHP defined in (\ref{ahp}) satisfies
	\begin{equation}\label{def2}	{\rm AHP}_{n}(\omega_{\nu},\alpha,\psi)=n^{-1}|z_{n}(\omega_{\nu},\alpha,\psi)|^{2}.
	\end{equation}	
	Together, (\ref{ahp}) and (\ref{def2}) show that the PG and AHP admit parallel Fourier and regression representations.

	\iffalse
	\noindent %{\it Remark (Normalization):}  
	{\it Remark 1.} In Sections \ref{sim} and \ref{APP}, we only focus on detecting periodicities and comparing serial dependence patterns in time series, and absolute spectral magnitude is of secondary interest. As a result, we normalize the periodograms so that the summation over $\omega$ equals 1 for each $(\alpha, \psi)$. %The purpose and implications of such normalization are discussed in Section \ref{CO}. 
	Additionally, we use the normalized frequency $f=\omega/2\pi \in [0,1/2]$, which represents the number of cycles per unit time, in all figures. %\emph{A caveat}: 
	Note that normalization alters the chi-square distribution and introduces dependence among ordinates; consequently, the confidence intervals developed in Section \ref{CIs} are valid only for the unnormalized AHP.
	\fi
	
	%The normalization of periodograms (sum-to-one at each $(\alpha,\psi)$) transforms the periodogram into a discrete spectral distribution that emphasizes relative power across frequencies. This is appropriate for detecting periodicities and comparing serial dependence patterns, but it removes absolute scale information and invalidates the chi-square-based CIs developed in Section~3.2 (which require the unnormalized AHP). Researchers who require both scale information and CIs should work with unnormalized periodograms.
	
	\subsection{Asymmetric Huber Spectrum}
	Analogous to the relationship between the PG and the power spectrum, we define the AHS via the autocovariance framework. The AHS is fundamentally the ordinary power spectrum of the asymmetrically truncated residual process $\{\dot{\rho}_{\alpha,\psi}(y_t - \mu({\alpha,\psi}))\}$, scaled by~$\eta^2(\alpha,\psi)$ to ensure that the asymptotic mean of the raw AHP coincides with the AHS. This perspective is parallel to: (i)~the quantile spectrum of \cite{li2012quantile}, which is a scaled version of the spectrum of the level-crossing (indicator) process $\{I(y_t\leq q_\alpha)\}$; and (ii)~the expectile spectrum of \cite{chen2026expectile}, which is a scaled version of the spectrum of the asymmetrically weighted process $\{\dot\rho_{\alpha}(y_t-\mu(\alpha))\}$.
	
	For intermediate values $0<\psi<\infty$, the process $\dot\rho_{\alpha,\psi}(y_t-\mu)$ has the following interpretation, which interpolates between the QP and EP extremes. Define the censored residual
	\[
	u_t(\alpha,\psi)
	:=
	\begin{cases}
		\alpha\bigl\{y_t-\mu(\alpha,\psi)\bigr\},
		& 0<y_t-\mu(\alpha,\psi)<\psi,\\
		\alpha\psi,
		& y_t-\mu(\alpha,\psi)\geq\psi,\\
		(1-\alpha)\bigl\{y_t-\mu(\alpha,\psi)\bigr\},
		& -\psi<y_t-\mu(\alpha,\psi)<0,\\
		-(1-\alpha)\psi,
		& y_t-\mu(\alpha,\psi)\leq-\psi.
	\end{cases}
	\]
	Then $\dot\rho_{\alpha,\psi}(y_t-\mu(\alpha,\psi)) \equiv u_t(\alpha,\psi)$, and the AHS is $\eta^2(\alpha,\psi)$ times the ordinary spectrum of $\{u_t(\alpha,\psi)\}$. As $\psi\to\infty$, the censoring disappears and $u_t(\alpha,\psi)$ becomes the asymmetrically weighted residual of ER, yielding the expectile spectrum. As $\psi\to0$, $u_t(\alpha,\psi)/\psi$ converges to the level-crossing indicator, yielding the quantile spectrum. For finite $\psi$, the AHS captures frequency-domain dependence among residuals that are linearly weighted within the central band $(-\psi,\psi)$ and truncated beyond it. The truncation bounds the influence of extreme observations, thereby providing robustness against outliers while retaining the smooth, differentiable structure of the quadratic region for central observations. This interpretation is particularly relevant in financial applications, where $\dot\rho_{\alpha,\psi}(y_t-\mu)$ can be viewed as a robustified version of the marginal contribution to an asymmetric risk measure that is sensitive to the asymmetry between gains and losses but insensitive to the exact magnitude of extreme events beyond a chosen threshold.
	
	Assume that the sequence $\{\dot{\rho}_{\alpha,\psi}(y_t - \mu({\alpha,\psi}))\}$ is stationary, and define the autocovariance function (ACF) of $\left\{ \dot{\rho}_{\alpha,\psi}(y_t - \mu(\alpha,\psi)) \right\}$ as
	\[
	\gamma(\tau, {\alpha,\psi}):={\rm Cov}\{\dot{\rho}_{\alpha,\psi}(y_{t+\tau} - \mu({\alpha,\psi})),\dot{\rho}_{\alpha,\psi}(y_t - \mu({\alpha,\psi}))\},
	\qquad\tau\in\mathbb Z.
	\]
	Assume that $\sum_{\tau=-\infty}^{\infty}|\gamma(\tau,{\alpha,\psi})| < \infty$ and let $\omega \in [0, \pi]$. Then the AHS is defined as
	\begin{equation}\label{es}
		g(\omega ,{\alpha,\psi}): = \eta^2({\alpha,\psi})\sum_{\tau=-\infty}^{\infty} \gamma(\tau,{\alpha,\psi})\exp(-i \omega \tau)
		= \eta^2({\alpha,\psi})\sum_{\tau=-\infty}^{\infty} \gamma(\tau,{\alpha,\psi})\cos(\omega\tau),
	\end{equation}
	where the second equality follows from the evenness of the autocovariance function,
	$\gamma(-\tau,\alpha,\psi)=\gamma(\tau,\alpha,\psi)$. For $\omega=0$, the AHS reduces to $g(0,\alpha,\psi)=\eta^2(\alpha,\psi)\sum_{\tau}\gamma(\tau,\alpha,\psi)$, which is $\eta^2(\alpha,\psi)$ times the long-run variance of the $\dot\rho_{\alpha,\psi}$ process. For $\omega=\pi$, the analogous definition holds with $\cos(\pi\tau)=(-1)^\tau$.	
	Assume that $\mathrm{E}\{\ddot\rho_{\alpha,\psi}
	(y_t-\mu(\alpha,\psi))\}>0$. Then, the scaling factor $\eta(\alpha,\psi)>0$ is defined through its reciprocal
	\[
	\begin{aligned}
		\eta^{-1}(\alpha,\psi) & :=  \mathrm{E}\left\{\ddot{\rho}_{\alpha,\psi}(y_t-\mu(\alpha,\psi))\right\}  \\
		&= \alpha\, \mathrm{Pr}\bigl(0 < y_t-\mu(\alpha,\psi) < \psi\bigr) \;+\; (1 - \alpha)\, \mathrm{Pr}\bigl(-\psi < y_t-\mu(\alpha,\psi) < 0\bigr) \\
		&= \alpha \,\bigl\{F(\psi+\mu(\alpha,\psi))-F(\mu(\alpha,\psi))\bigr\} \;+\; (1-\alpha)\,\bigl\{F(\mu(\alpha,\psi))-F(\mu(\alpha,\psi)-\psi)\bigr\}, 
	\end{aligned}
	\]
	which depends solely on the marginal distribution function \(F(\cdot)\) and the parameters ($\alpha, \psi$). The equality follows from
	\[
	\mathrm{E}\{\ddot\rho_{\alpha,\psi}(U_t)\}
	= \int_{-\infty}^{\infty} \ddot\rho_{\alpha,\psi}(u)\,f_U(u)\,du,
	\]
	where the second derivative $\ddot{\rho}_{\alpha,\psi}(u) = \partial \dot{\rho}_{\alpha,\psi}(u)/\partial u$ exists almost everywhere with respect to the Lebesgue measure and is given by
	\begin{equation}\label{psia}
		\ddot{\rho}_{\alpha,\psi}(u)
		=
		\alpha\,\mathbf{1}(0<u<\psi)
		+
		(1-\alpha)\,\mathbf{1}(-\psi<u<0),
		\qquad \text{a.e. }u\in\mathbb{R}.
	\end{equation}
	The values of $\ddot{\rho}_{\alpha,\psi}$ at the three kink points
	$u\in\{-\psi,0,\psi\}$ are unspecified, but under condition~(C1) in Section~3.1
	the residual $U_t=y_t-\mu(\alpha,\psi)$ has a density $f_U$, so
	these three singletons have zero Lebesgue measure and do not affect
	the expectation. The condition $\eta^{-1}(\alpha,\psi)>0$
	ensures that the population objective function
	$\mu\mapsto\mathrm{E}\{\rho_{\alpha,\psi}(y_t-\mu)\}$ is strictly
	convex at its minimizer, which in turn guarantees uniqueness of
	$\mu(\alpha,\psi)$ and the validity of the normal equation~\eqref{ne}.
	
	\section{Asymptotic Analysis}\label{AS}
	\subsection{Large Sample Theory}
	We develop a large-sample asymptotic analysis that reveals an association between the AHP and the AHS. Throughout this section, $(\alpha,\psi)$ is fixed. %For fixed $\alpha$ and $\psi$, 
	Let $\{y_t\}$ be a stationary time series with cumulative distribution function $F(\cdot)$. We consider the joint asymptotic distribution of the AHP ordinates computed at $q$ distinct frequencies, denoted by $ \{\omega_{\nu_{j}}: j=1,\ldots,q\}$. Let $\mathbf{x}_{jnt} =[ 1 , \cos(\omega_{\nu_j} t) , \sin(\omega_{\nu_j} t) ]^{\top}$ for $t = 1, \ldots, n$ and $j=1,\ldots, q.$
	Replacing $\mathbf{x}_{nt}$ with $\mathbf{x}_{jnt}$ in (\ref{er}) yields $\hat{\bm{\beta}}_{jn}(\alpha,\psi)$ and ${\rm AHP}_{n}(\omega_{\nu_{j}},\alpha,\psi)$. We define 
	\begin{equation}\label{zetanj}
		\boldsymbol{\zeta}_{jn}(\alpha,\psi):=\frac{1}{\sqrt n}\sum_{t=1}^{n}\dot{\rho}_{\alpha,\psi}(y_{t}-\mu(\alpha,\psi))\mathbf{x}_{jnt}.
	\end{equation}
	
	Let the following conditions be satisfied for fixed $\alpha$ and $\psi$:
	\begin{itemize}
		\item[C1.] Let
		$
		U_t=y_t-\mu(\alpha,\psi).
		$
		There exists some $\epsilon>0$ such that $U_t$ has a marginal density
		$f_U$ on
		$
		\mathcal N_\epsilon
		=
		\bigcup_{a\in\{-\psi,0,\psi\}}
		(a-\epsilon,a+\epsilon),
		$
		and
		$
		\sup_{u\in\mathcal N_\epsilon}f_U(u)<\infty.
		$
		Moreover, the expected second derivative is strictly positive:
		$
		\eta^{-1}(\alpha,\psi)=\mathrm{E}\left\{
		\ddot\rho_{\alpha,\psi}(U_t)
		\right\}>0.
		$
		
		\item[C2.]  The process $\{y_{t}\}$ has the strong mixing property with mixing numbers $a_{\tau}$ ($\tau=1,2,\ldots$) satisfying $a_{\tau}\to 0$ as $\tau\to\infty$. 
		
		\item[C3.] The process $\{\dot{\rho}_{\alpha,\psi}(y_t - \mu(\alpha,\psi))\}$ is (strictly) stationary with zero mean and absolutely summable ACF defined as $\text{Cov}\{\dot{\rho}_{\alpha,\psi}(y_t - \mu(\alpha,\psi)),\dot{\rho}_{\alpha,\psi}(y_s - \mu(\alpha,\psi))\} = \gamma(t-s,\alpha,\psi)$. Under (C2), the strict stationarity of $\{y_t\}$ implies the strict stationarity of $\{\dot\rho_{\alpha,\psi}(y_t-\mu(\alpha,\psi))\}$ because $\mu(\alpha,\psi)$ is a time-invariant functional of the marginal distribution. 
		
		\item[C4.]A joint central limit theorem holds for the score vectors: let $\mathbf{V}_{jl}(\alpha,\psi)$ denote the asymptotic covariance matrix between $\boldsymbol{\zeta}_{jn}(\alpha,\psi)$ and $\boldsymbol{\zeta}_{ln}(\alpha,\psi)$ defined in~\eqref{zetanj}; then
		\[
		\operatorname{vec}[\boldsymbol{\zeta}_{jn}(\alpha,\psi)]_{j=1}^q:=[\boldsymbol{\zeta}_{1n}^{\top},\dots,\boldsymbol{\zeta}_{qn}^{\top}]^{\top}\xrightarrow{D}N(\mathbf{0},[\mathbf{V}_{jl}(\alpha,\psi)]_{j,l=1}^q),\quad \text{as } n\to\infty.
		\]
		Explicit expressions for $\mathbf{V}_{jl}(\alpha,\psi)$ are derived in the proof of Theorem~1 (Appendix~A.2).
		
	\end{itemize}
	
	Equipped with these concepts, we establish the asymptotic behavior of the ${\rm AHP}_{n}(\omega_{\nu_j},\alpha,\psi)$.
	\begin{theorem} %(Asymmetric Huber Periodogram).
		For fixed \(q\ge1\) and \(0<\lambda_{1}<\cdots<\lambda_{q}<\pi\), let \(\omega_{\nu_{1}},\ldots,\omega_{\nu_{q}}\) be Fourier frequencies satisfying \(\omega_{\nu_{j}}\to\lambda_{j}\) as \(n\to\infty\) for \(j=1,\ldots,q\). If conditions (C1)--(C4) hold, then
		\begin{equation}\label{the2}
			\{{\rm AHP}_{n}(\omega_{\nu_{1}},\alpha,\psi),\ldots,{\rm AHP}_{n}(\omega_{\nu_{q}},\alpha,\psi)\}\xrightarrow{D}\{g(\lambda_{1},\alpha,\psi)(1/2)\chi^{2}_{2,1},\ldots,g(\lambda_{q},\alpha,\psi)(1/2)\chi^{2}_{2,q}\},
		\end{equation} 
		where \(\chi^{2}_{2,1},\ldots,\chi^{2}_{2,q}\) are independent standard chi-square random variables with two degrees of freedom and $\xrightarrow{D}$ denotes convergence in distribution.
	\end{theorem}
	\noindent
	{\it Proof.} See Appendix.
	
	Theorem 1 shows that the AHP exhibits a scaled chi-square distribution with two degrees of freedom, similar to the behavior of the PG \citep{brockwell1991time}.
	
	To obtain convergence of its first two moments, we additionally assume
	that, for some $\delta>0$,
	\begin{equation}\label{moment_condition}
		\sup_{n\geq 1}
		\mathrm{E}\left[
		\left\{
		\operatorname{AHP}_n
		(\omega_{\nu_n},\alpha,\psi)
		\right\}^{2+\delta}
		\right]
		<\infty
	\end{equation}
	whenever $\omega_{\nu_n}\to\lambda\in(0,\pi)$. This condition implies
	the uniform integrability of
	$\{
	\operatorname{AHP}_n^2
	(\omega_{\nu_n},\alpha,\psi)
	\}$.
	Consequently, Theorem~1 and the convergence-of-moments theorem yield
	$$
	\mathrm{E}\left\{
	\operatorname{AHP}_n
	(\omega_{\nu_n},\alpha,\psi)
	\right\}
	\longrightarrow
	g(\lambda,\alpha,\psi),
	$$
	and
	$$
	\operatorname{Var}\left\{
	\operatorname{AHP}_n
	(\omega_{\nu_n},\alpha,\psi)
	\right\}
	\longrightarrow
	g^2(\lambda,\alpha,\psi).
	$$
	Therefore, under the additional moment condition \eqref{moment_condition}, the raw AHP is asymptotically unbiased for the AHS but is not a consistent estimator, since its asymptotic variance does not vanish.

	Let the smoothed estimator be
	\begin{equation}\label{smoahp}
		\hat{g}(\omega_{\nu}, \alpha,\psi) 
		= \sum_{s=-M_n}^{M_n} W_{n,s}\,
		{\rm AHP}_{n}(\omega_{\nu+s},\alpha,\psi),
	\end{equation}
	where $\omega_{\nu+M_{n}}, \omega_{\nu-M_{n}} \in [0, \pi]$  and $W_{n,s}\geq 0$ are weights satisfying $\sum_{s=-M_n}^{M_n} W_{n,s} = 1$. In addition to conditions (C1)--(C4), we assume
	\begin{itemize}
		\item[C5.]  The function $g(\omega,\alpha,\psi)$ is continuous at $\lambda\in(0,\pi)$.
		\item[C6.] The smoothing span satisfies $M_n\to\infty$, $M_n/n\to0$,
		and $\omega_\nu\to\lambda$ as $n\to\infty$.
		\item[C7.] The weights $\{W_{n,s}: |s|\leq M_n\}$ are nonnegative and satisfy
		\[
		\sum_{s=-M_n}^{M_n} W_{n,s}=1,
		\qquad
		\max_{|s|\leq M_n}W_{n,s}=O(M_n^{-1}).
		\]
		
		\item[C8.] Let
		$
		A_{n,s}
		=
		\operatorname{AHP}_n
		(\omega_{\nu+s},\alpha,\psi).
		$
		The weighted cross-covariances within the smoothing window satisfy
		\[
		\sum_{\substack{|s|\leq M_n,\ |r|\leq M_n\\s\neq r}}
		W_{n,s}W_{n,r}
		\left|
		\operatorname{Cov}(A_{n,s},A_{n,r})
		\right|
		\longrightarrow0.
		\]
	\end{itemize}
	Then, we establish the following theorem regarding the smoothed estimator $\hat{g}(\omega_{\nu}, \alpha,\psi)$.
	\begin{theorem} %(Smoothed Asymmetric Huber Periodogram). 
		If conditions (C1)--(C8) and \eqref{moment_condition} hold, then
		\begin{equation}\label{p3}
			\hat{g}(\omega_{\nu}, \alpha,\psi)  \;\overset{p}{\longrightarrow}\; g(\lambda, \alpha,\psi).
		\end{equation}
	\end{theorem}
	\noindent
	{\it Proof.} See Appendix.
	
	Theorem 2 indicates that the smoothed AHP provides a consistent estimator of the AHS.  This implies that the AHS can be estimated directly by smoothing the raw AHP, thereby removing the need to estimate the scaling factor $\eta^2(\alpha,\psi)$. 
	
	\noindent %{\it Remark (Convergence rate and asymptotic normality).}
	{\it Remark 1.} Theorem~2 establishes the consistency of the smoothed AHP but does
	not provide an explicit convergence rate or a limiting distribution.
	Conditions~(C1)--(C8) are sufficient for the bias and variance to
	vanish, but they do not specify their rates. In particular,
	condition~(C8) only requires the weighted off-diagonal covariance
	term to be $o(1)$, while condition~(C5) imposes continuity rather than
	a quantitative smoothness condition on the AHS.
	Obtaining an explicit convergence rate would require stronger uniform
	conditions on the finite-sample bias and cross-frequency covariances
	of the raw AHP, together with additional smoothness of
	$g(\cdot,\alpha,\psi)$. Similarly, a formal central limit theorem for
	the smoothed AHP would require stronger fourth-order dependence
	conditions and a triangular-array limit theory for the ordinates in
	a shrinking frequency neighborhood. These developments are beyond
	the scope of the present paper. Theorem~1 nevertheless directly
	supports inference based on the raw AHP, while inference based on the
	smoothed AHP below additionally uses a local-independence
	approximation.
	
	\subsection{Confidence Intervals}\label{CIs}
	This subsection develops pointwise confidence intervals for the AHS
	based on Theorem~1 and an additional local-independence approximation
	for the smoothed AHP. Throughout, $(\alpha,\psi)$ is fixed and $\omega_\nu\to\lambda\in(0,\pi)$.
	
	Under the conditions of Theorem~1, for a fixed sequence of interior
	Fourier frequencies satisfying $\omega_\nu\to\lambda$,
	\[
	\frac{2\operatorname{AHP}_n(\omega_\nu,\alpha,\psi)}
	{g(\lambda,\alpha,\psi)}
	\xrightarrow{d}\chi_2^2.
	\] 
	This distributional approximation yields a pointwise CI for the AHS. Let $\chi^2_{d,\gamma}$ denote the $\gamma$-quantile of the chi-square distribution with $d$ degrees of freedom. An approximate $100(1-\gamma)\%$ pointwise CI for $g(\lambda,\alpha,\psi)$ based on the raw AHP is
	\begin{equation}
		\label{CI1}
		\mathrm{CI}_{1-\gamma}^{(1)}(\lambda;\alpha,\psi)
		=
		\left[
		\frac{2\,{\rm AHP}_n(\omega_\nu,\alpha,\psi)}{\chi^2_{2,\,1-\gamma/2}},
		\quad
		\frac{2\,{\rm AHP}_n(\omega_\nu,\alpha,\psi)}{\chi^2_{2,\,\gamma/2}}
		\right].
	\end{equation}
	
	Next, consider the smoothed estimator in \eqref{smoahp}. For equal weights,
	let
	$$
	W_{n,s}=\frac{1}{L_n},
	\qquad
	L_n=2M_n+1,
	\qquad
	s=-M_n,\ldots,M_n.
	$$
	Under the additional approximation that the raw AHP ordinates within
	the smoothing window are asymptotically independent and have the
	common local scale $g(\lambda,\alpha,\psi)$, we have
	\[
	\frac{2\,\operatorname{AHP}_n
		(\omega_{\nu+s},\alpha,\psi)}
	{g(\lambda,\alpha,\psi)}
	\overset{a}{\sim}\chi_2^2,
	\qquad s=-M_n,\ldots,M_n.
	\]
	Consequently,
	\[
	\frac{2L_n\hat g
		(\omega_\nu,\alpha,\psi)}
	{g(\lambda,\alpha,\psi)}
	\overset{a}{\sim}
	\chi_{2L_n}^2.
	\]
	
	An approximate $100(1-\gamma)\%$ pointwise CI for $g(\lambda,\alpha,\psi)$ based on the smoothed AHP %estimator $\hat{g}(\omega_{\nu}, \alpha,\psi)$ 
	is
	\begin{equation}
		\label{CI2}
		\mathrm{CI}_{1-\gamma}^{(2)}(\lambda;\alpha,\psi)
		=
		\left[
		\frac{2L_n\,\hat g(\omega_\nu,\alpha,\psi)}{\chi^2_{2L_n,\,1-\gamma/2}},
		\quad
		\frac{2L_n\,\hat g(\omega_\nu,\alpha,\psi)}{\chi^2_{2L_n,\,\gamma/2}}
		\right].
	\end{equation}
	
	Note that intervals \eqref{CI1} and \eqref{CI2} are pointwise in frequency. Simultaneous bands may be constructed either by applying multiplicity corrections over a frequency grid or by using frequency-domain bootstrap methods. In view of the shape of the chi-square distribution, smoothing effectively increases the degrees of freedom from $2$ to $2L_n$, which substantially shortens the  CI and yields a more stable estimator of the AHS.

	In many applications, spectral features may be more clearly visualized on the logarithmic scale. The log transformation serves as a variance-stabilizing transformation. This is particularly useful when peaks of interest are much smaller than some dominant components. The CIs for $\log g(\omega,\alpha,\psi)$ are
	\begin{equation}\label{log1}
		\left[
		\log {\rm AHP}_n(\omega,\alpha,\psi)+\log 2-\log \chi^2_{2,1-\frac{\gamma}{2}},
		\ \ 
		\log {\rm AHP}_n(\omega,\alpha,\psi)+\log 2-\log \chi^2_{2,\frac{\gamma}{2}}
		\right],
	\end{equation}
	and
	\begin{equation}
		\label{log2}
		\left[
		\log \hat g(\omega,\alpha,\psi) + \log(2L_n) - \log \chi^2_{2L_n,1-\frac{\gamma}{2}},
		\ \ 
		\log \hat g(\omega,\alpha,\psi) + \log(2L_n) - \log \chi^2_{2L_n,\frac{\gamma}{2}}
		\right].
	\end{equation}
	
	For unequal smoothing weights, the weighted sum of the raw AHP ordinates does not have an exact chi-square distribution. We therefore use a Satterthwaite moment-matching approximation and define the effective number of averaged ordinates as $L_{\mathrm{eff}}=(\sum_{s=-M_n}^{M_n}W_{n,s}^2)^{-1}$. Then, $L_n$ in \eqref{CI2} and \eqref{log2} are replaced by $L_{\mathrm{eff}}$.
	
	\noindent %{\it Remark (Normalization):}
	{\it Remark 2.} In Sections \ref{sim} and \ref{APP}, we only focus on detecting periodicities and comparing serial dependence patterns in time series, and absolute spectral magnitude is of secondary interest. As a result, we normalize the periodograms so that the summation over Fourier frequencies $\omega$ equals 1 for each fixed $(\alpha, \psi)$. %In Sections~\ref{sim} and~\ref{APP}, periodograms are normalized so that the sum over Fourier frequencies equals~$1$ for each fixed $(\alpha,\psi)$. 
	This transformation converts the periodogram into a discrete probability distribution over frequencies, emphasizing relative spectral shape rather than absolute magnitude. However, normalization introduces a random denominator common to all ordinates, which alters the chi-square distribution and induces dependence among the ordinates. Consequently, the confidence intervals constructed in this subsection apply only to the unnormalized (raw or smoothed) AHP, and should not be applied directly to the normalized version. The spectral shape information preserved by normalization suffices for periodicity detection (Section~\ref{fish}) and visual comparison, but scale-dependent inference requires the unnormalized periodogram.
	
	\subsection{Fisher-Type Test}\label{fish}
	
	We next develop a Fisher-type test for detecting a dominant periodic or oscillatory component based on the AHP. %Throughout this subsection, $(\alpha,\psi)$ is fixed. 
	Let $\mathcal{F}_n=\{\omega_{\nu_{1,n}},\ldots,\omega_{\nu_{q,n}}\}\subset (0,\pi)$ be a prespecified collection of $q\geq 2$ distinct, nonredundant Fourier frequencies, where $q$ is fixed and $\omega_{\nu_{j,n}}\longrightarrow\lambda_j$ as $n\to\infty$ for $j=1,\ldots,q$, with $0<\lambda_1<\cdots<\lambda_q<\pi$. Thus, both the zero frequency and the Nyquist frequency are excluded.
	
	The classical Fisher test for the ordinary periodogram \citep{fisher1929tests} is formulated to detect a dominant sinusoidal component against a flat-spectrum null hypothesis. In the present setting, we consider testing the null hypothesis 
	\[
	H_0:
	\quad
	g(\lambda_1,\alpha,\psi)
	=
	\cdots
	=
	g(\lambda_q,\alpha,\psi)
	=
	g_0(\alpha,\psi)
	\]
	against the alternative hypothesis
	\[
	H_1:
	\quad
	g(\lambda_j,\alpha,\psi)>g_0(\alpha,\psi)
	\quad
	\text{for at least one }
	j\in\{1,\ldots,q\}.
	\]
	Thus, under $H_0$, the AHS is flat over the prespecified frequency set,
	whereas $H_1$ indicates the presence of a dominant periodic or oscillatory component at one or more candidate frequencies.
	
	We define the AHP-based Fisher statistic as
	\[
	G_{n,\mathrm{AHP}}(\alpha,\psi)
	=
	\frac{
		\displaystyle
		\max_{1\leq j\leq q}
		\mathrm{AHP}_n
		\bigl(\omega_{\nu_{j,n}},\alpha,\psi\bigr)
	}{
		\displaystyle
		\sum_{j=1}^{q}
		\mathrm{AHP}_n
		\bigl(\omega_{\nu_{j,n}},\alpha,\psi\bigr)
	}.
	\]
	A large value of $G_{n,\mathrm{AHP}}(\alpha,\psi)$ indicates that a substantial proportion of the spectral contribution is concentrated at a single candidate frequency, providing evidence against $H_0$.
	
	Under $H_0$, Theorem~1 implies that, for fixed $q$,
	\[
	\left(
	\frac{
		\mathrm{AHP}_n
		\bigl(\omega_{\nu_{1,n}},\alpha,\psi\bigr)
	}{
		g_0(\alpha,\psi)
	},
	\ldots,
	\frac{
		\mathrm{AHP}_n
		\bigl(\omega_{\nu_{q,n}},\alpha,\psi\bigr)
	}{
		g_0(\alpha,\psi)
	}
	\right)
	\xrightarrow{d}
	(E_1,\ldots,E_q),
	\]
	where $E_1,\ldots,E_q$ are independent standard exponential random
	variables, since $\chi_2^2/2\sim\mathrm{Exp}(1)$. Because the common
	scale factor $g_0(\alpha,\psi)$ cancels from the ratio, the continuous
	mapping theorem gives
	\[
	G_{n,\mathrm{AHP}}(\alpha,\psi)
	\xrightarrow{d}
	G_q
	:=
	\frac{
		\max_{1\leq j\leq q}E_j
	}{
		\sum_{j=1}^{q}E_j
	}.
	\]
	Therefore, for $0<x<1$,
	\[
	\Pr(G_q>x)
	=
	\sum_{k=1}^{\min\{q,\lfloor 1/x\rfloor\}}
	(-1)^{k-1}
	\binom{q}{k}
	(1-kx)^{q-1}.
	\]
	
	For an observed value
	$
	g_{\mathrm{obs}}
	=
	G_{n,\mathrm{AHP}}(\alpha,\psi),
	$
	the corresponding asymptotic $p$-value is
	\[
	p_{\mathrm{AHP}}
	=
	\sum_{k=1}^{\min\{q,\lfloor 1/g_{\mathrm{obs}}\rfloor\}}
	(-1)^{k-1}
	\binom{q}{k}
	(1-kg_{\mathrm{obs}})^{q-1}.
	\]
	The null hypothesis $H_0$ is rejected at significance level $\gamma\in(0,1/2]$ whenever
	$p_{\mathrm{AHP}}<\gamma$.
	
	%\noindent
	%{\it Remark (Testing over all Fourier frequencies).}
	%The test described above is designed for a pre-specified collection of $q$ candidate frequencies, which is appropriate when prior knowledge (e.g., from subject-matter theory) suggests specific periodicities. The classical Fisher test for the PG is applied over \emph{all} $Q=\lfloor(n-1)/2\rfloor$ Fourier frequencies in $(0,\pi)$. Extending our test to this setting is complicated by two factors: (i)~as $q=Q\to\infty$ with $n$, the $q$ ordinates are no longer asymptotically independent (neighboring Fourier frequencies have non-negligible covariance), so the limiting distribution of the ratio statistic is not a simple function of independent exponentials; (ii)~the standard large-$q$ approximation for the PG \citep{fisher1929tests} relies on the exact independence of periodogram ordinates under Gaussian white noise, which does not carry over to the AHP. In the simulation studies (Section~4.2), we apply the test to a single candidate frequency $\omega=0.171\times2\pi$, which is the known spectral peak frequency of the AR(2) process. For applications without a specific candidate frequency, we recommend using the test with all Fourier frequencies and calibrating the null distribution via a frequency-domain bootstrap \citep{franke1986bootstrap,dahlhaus1988empirical}.
	
	\section{Simulation Studies}\label{sim}
	In this section, we present experiments to evaluate the effectiveness of the AHP in detecting hidden periodicities and its robustness against outliers, along with comparisons to other types of periodograms.

	\subsection{Hidden Periodicity Detection}
	Suppose that $\{x_t\}$ is an AR(2) model:
	\begin{equation}\label{ar2}
		x_t = \phi_1 x_{t-1} + \phi_2 x_{t-2} + \epsilon_t,
	\end{equation}
	and 
	\begin{equation}\label{hidden}
		y_t = \{b_0 + b_1\cos(\omega_{\nu_0} t) + b_2\sin(\omega_{\nu_1} t)\} x_t, 
	\end{equation} 
	where $b_0=1, b_1=0.9, b_2=1$, $\phi_1 = 2r \cos(\omega_{\nu_c}), \phi_2 = -r^2$ ($r=0.6)$, and $\{\epsilon_t\}$ is the standard Gaussian white noise.
	Under this setting, $\{y_t\}$ is obtained by multiplying the carrier process $\{x_t\}$ (an AR(2) with its own spectral peak near $\omega_{\nu_c}$) by a slowly varying amplitude modulator $\{b_0 + b_1\cos(\omega_{\nu_0} t) + b_2\sin(\omega_{\nu_1} t)\}$ (which contains the hidden frequencies $\omega_{\nu_0} = 0.09 \times 2 \pi$ and $ \omega_{\nu_1} = 0.12 \times 2\pi$). The sample size is set as $n=200$.

\begin{figure}[H]
		\centering
		\includegraphics[width=\textwidth]{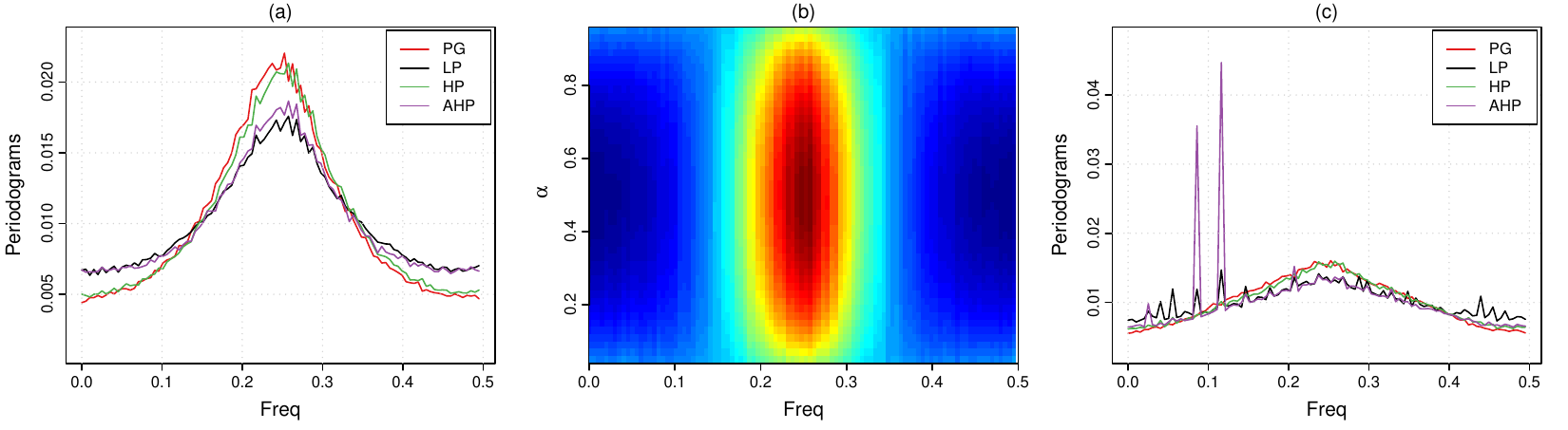}
		\caption{Left: averaged PG, LP, HP, and AHP ($\alpha = 0.8, \psi=1.345 \hat{\sigma}_{x_t}$) of model (\ref{ar2}). Middle: averaged AHP $(\alpha=0.05, 0.07,\cdots, 0.95$, $\psi=1.345 \hat{\sigma}_{x_t}$) of model (\ref{ar2}).  Right: PG, LP, HP, and AHP ($\alpha=0.8$ and $\psi=1.345 \hat{\sigma}_{y_t}$) of model (\ref{hidden}). The number of realizations is $5{,}000$ and 1.345 is suggested by \cite{huber1964robust}.} 
		\label{smo}
	\end{figure}

	We first present the averaged periodograms from $5{,}000$ simulations of model (\ref{ar2}) with PG, LP, and HP for comparison. As shown in Fig.~\ref{smo} (a), all types of periodograms exhibit similar bell-shaped patterns, with a maximum value at $\omega_{\nu_c} = 0.25\times 2\pi$. Fig.~\ref{smo} (b) further shows that this pattern persists across the considered values of $\alpha$. These results can be directly explained by the ordinary spectral analysis.  Specifically, the characteristic polynomial of model (\ref{ar2}) is $\phi(z)=1-\phi_1 z-\phi_2 z^2$, where the roots $z_1$ and $z_2$ are complex conjugates. The AR coefficients $\phi_1$ and $\phi_2$ determine the location of the spectral peak: 
	\[
	\omega = \arccos \frac{\phi_1}{2 \sqrt{-\phi_2}}  = \omega_{\nu_c},
	\] and as $r \rightarrow 1^-$, the peak becomes sharper, where $r\in(0,1)$ is a damping parameter; equivalently, the roots of
	the AR characteristic polynomial have modulus $1/r$.
	
	For model (\ref{hidden}), as illustrated in Fig.~\ref{smo} (c), the AHP successfully detects the hidden periodicities, resulting in large spikes at $\omega_{\nu_0} = 0.09\times 2\pi$ and $\omega_{\nu_1}=0.12 \times 2 \pi$, whereas the other periodograms retain bell-shaped patterns similar to those observed for model \eqref{ar2} and fail to reveal the two hidden frequencies. We observe small spikes at other frequencies in the LP, caused by spectral leakage. This phenomenon is explained by Theorem 5 in \cite{li2008laplace} and Theorem 2 in \cite{li2012quantile}, and can also occur in the AHP.
	
\begin{figure}[ht]
		\centering
		%\subfigcapskip = -0.3cm
		\includegraphics[width=0.7\textwidth]{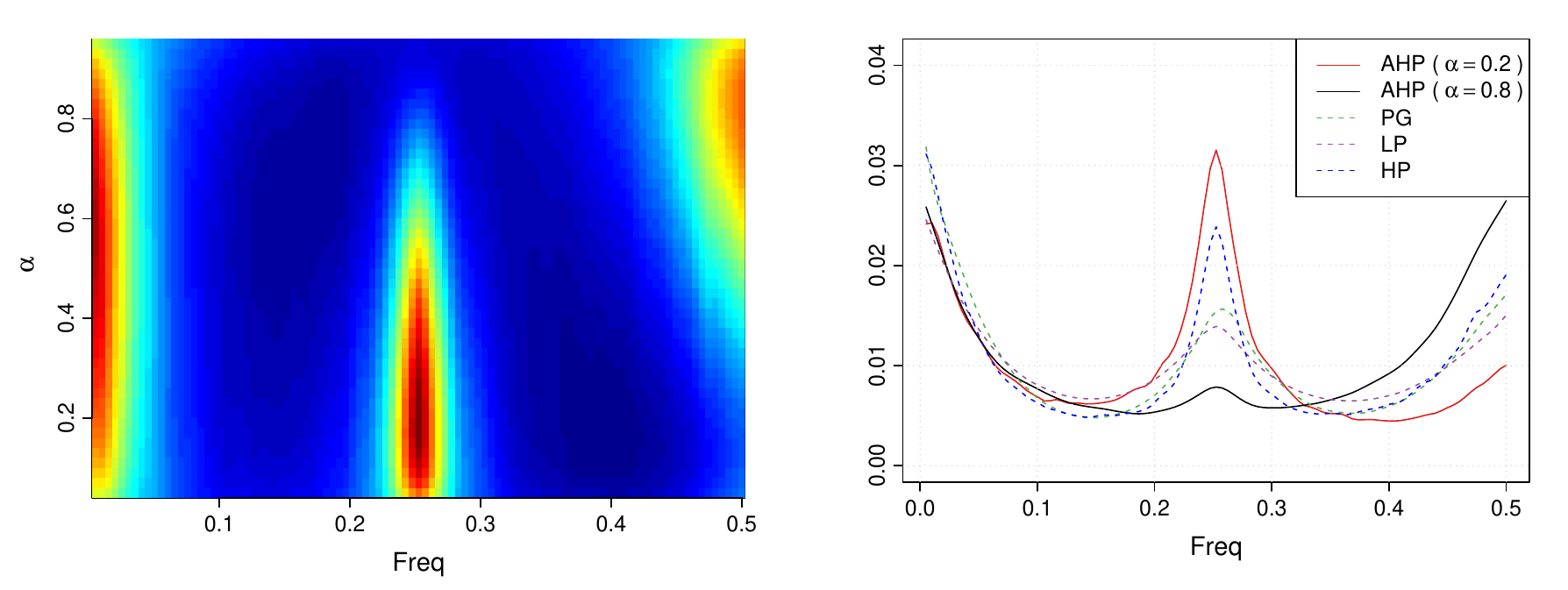}
		\caption{Left: AHP ($\alpha=0.05, 0.07,\cdots, 0.95$ and $\psi = 1.345\hat{\sigma}_{y_t}$). Right:  AHP ($\alpha=0.2,0.8$ and $\psi = 1.345\hat{\sigma}_{y_t}$), PG, LP, and HP. Results are based on the average of 5{,}000 smoothed periodograms.}
		\label{fig-asy}
	\end{figure}

	The AHPs of models (\ref{ar2}) and (\ref{hidden}) are symmetric across $\alpha$. To demonstrate the capability of the AHP in addressing more complex spectral features, we construct a nonlinear mixture model $\{y_t\}$, with an AHP intentionally asymmetric across $\alpha$:
	\begin{equation}\label{eq-mix}
		y_t  =   W_2(z_t) \, z_t + \{1-W_2(z_t)\} \, x_{t,3},	
	\end{equation}
	where
	$
	z_t  =   W_1(x_{t,1}) \, x_{t,1} + \{1-W_1(x_{t,1})\} \, x_{t,2}.
	$
	The components $\{x_{t,1}\}$, $\{x_{t,2}\}$, and $\{x_{t,3}\}$ are independent Gaussian AR processes given by $x_{t,1}  =  0.8 x_{t-1,1} + w_{t,1},$
	$x_{t,2}  =  -0.75 x_{t-1,2} + w_{t,2},$ and
	$x_{t,3} =  -0.81 x_{t-2,3} + w_{t,3},$
	where $w_{t,1}$, $w_{t,2}$, and $w_{t,3}$ are independent standard Gaussian white noise. From the perspective of traditional spectral analysis, the series $\{x_{t,1}\}$ has a lowpass spectrum, $\{x_{t,2}\}$ has a highpass spectrum, and $\{x_{t,3}\}$ has a bandpass spectrum around frequency $0.25 \times 2 \pi$. The mixing functions $W_1(x)$ and $W_2(x)$ are defined as 
	\[
	W_1(x) =
	\begin{cases}
		0.75, & x < -0.8, \\
		-\tfrac{13}{32}x + 0.425, & -0.8 \le x \le 0.8, \\
		0.1, & x > 0.8 ,
	\end{cases}
	\qquad
	\text{and}
	\qquad
	W_2(x) =
	\begin{cases}
		0.5, & x < -0.4, \\
		-1.25x, & -0.4 \le x \le 0, \\
		0, & x > 0 .
	\end{cases}
	\]
	Fig.~\ref{fig-asy} shows the averaged smoothed periodograms of model (\ref{eq-mix}) from 5{,}000 simulations, illustrating that the PG, LP, and HP each provide only a single spectral view and therefore fail to reveal how the spectral structure varies across the asymmetry level. In contrast, the AHP characterizes this variation over $\alpha \in (0, 1)$.
	
	\subsection{Robustness Evaluation Using Fisher's Test}\label{reuft}
\begin{figure}[ht]
		\centering
		\includegraphics[width=\textwidth]{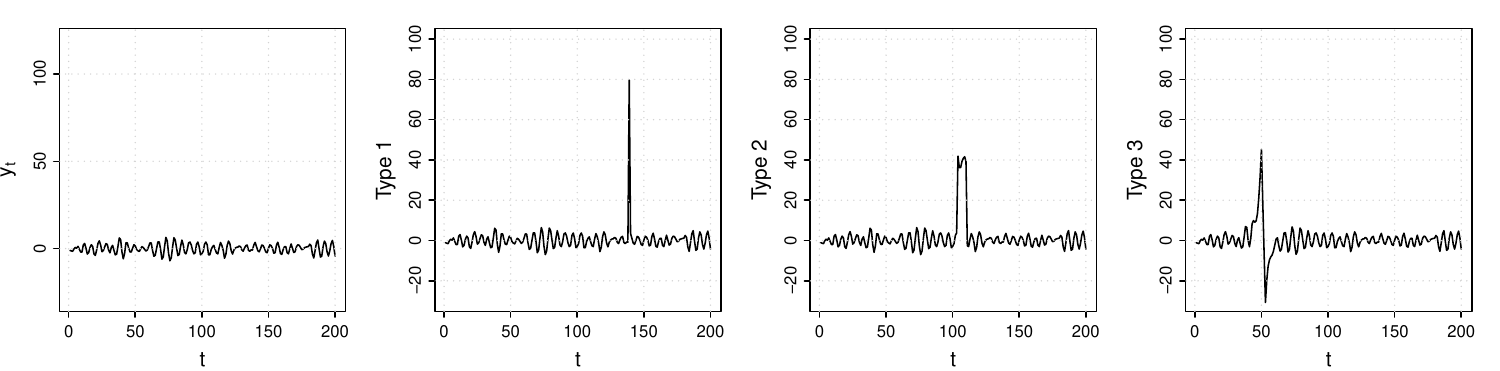}
		\caption{Time series and outlier-contaminated signals. Left to right: $y_t$, $y_{1,t}$, $y_{2,t}$, $y_{3,t}$. The sample size is $n=200$ and $(c_1,c_2,c_3) = (30, 15, 15)$ times the sample standard deviation of the time series.}
		\label{out}
	\end{figure}

	We apply Fisher's test to evaluate the robustness of the AHP.
	Suppose that $\{y_t\}$ is an AR(2) process with
	\begin{equation}\label{yt}
		y_t = 0.9y_{t-1}-0.9y_{t-2} + \epsilon_t,
	\end{equation}
	which has a single spectral peak at $0.171 \times 2\pi$.
	To evaluate the robustness of the AHP, we introduce three contamination models:
	\begin{itemize}
		\item Type 1 (single-point outlier): add a constant of magnitude $c_1$ to $\{y_t\}$ at a single time point $t^*$:
		\begin{equation}\label{type1}
			y_{1,t} =  
			\begin{cases} 
				y_t, & \text{if } t \ne t^* ,\\
				y_t + c_1, & \text{if } t = t^*, 
			\end{cases}	
		\end{equation}
		where $t^*$ denotes a random integer in $[1, n]$.
		
		\item Type 2 (short-term burst): add a peak of magnitude $c_2$ to $\{y_t\}$ over the interval $[t^*,t^*+5]$:
		\begin{equation}\label{type2}
			y_{2,t} =  
			\begin{cases} 
				y_t, & \text{if } t \notin [t^*,t^*+5] ,\\
				y_t + c_2, & \text{if } t\in [t^*,t^*+5], 
			\end{cases}	
		\end{equation}
		where $t^*$ denotes a random integer in $[1, n-5]$.
		
		\item Type 3 (eyeblink artifact): an eyeblink artifact is added to $\{y_t\}$ to simulate contamination commonly observed in EEG data \citep{mansor2011integrating}: 
		\begin{equation}\label{type3}
			y_{3,t} =  
			\begin{cases} 
				y_t, & \text{if } t \notin [t^*,t^*+50], \\
				y_t + \text{eyeblk}(c_3), & \text{if } t\in [t^*,t^*+50], 
			\end{cases}	
		\end{equation}
		where $\text{eyeblk}(c_3)$ denotes the eyeblink artifact generated by two Gamma functions with eye-closure magnitude $c_3$ and eye-open magnitude $0.6c_3$, and $t^*$ denotes a random integer in $[1, n-50]$.
	\end{itemize}
	Fig.~\ref{out} shows the three types of outliers.
	
	For each of the models (\ref{yt})--(\ref{type3}), we generate $1{,}000$ time series and compute the AHP, EP, and PG. We set $\alpha = 0.6, 0.8$, and  $\psi$ = $(0.674, 0.935, 1.345)$ times the sample standard deviation of the time series. For an uncontaminated Gaussian marginal distribution, these constants approximately correspond to central probabilities $0.50$, $0.65$, and $0.82$, respectively. Both the contamination magnitudes and the threshold parameter are scaled by the sample standard deviation of the uncontaminated series. Then, we estimate the detection probability (PD), or equivalently the empirical power of Fisher’s test, at significance levels 0.01 and 0.05. Tables \ref{tb1}--\ref{tb3} report the PD values for models \eqref{yt}--\eqref{type3}, along with the differences. The difference is defined as the PD for the uncontaminated series minus the PD for the corresponding contaminated series.
	
\begin{figure}[ht]
		\centering
		\includegraphics[width=0.9\textwidth]{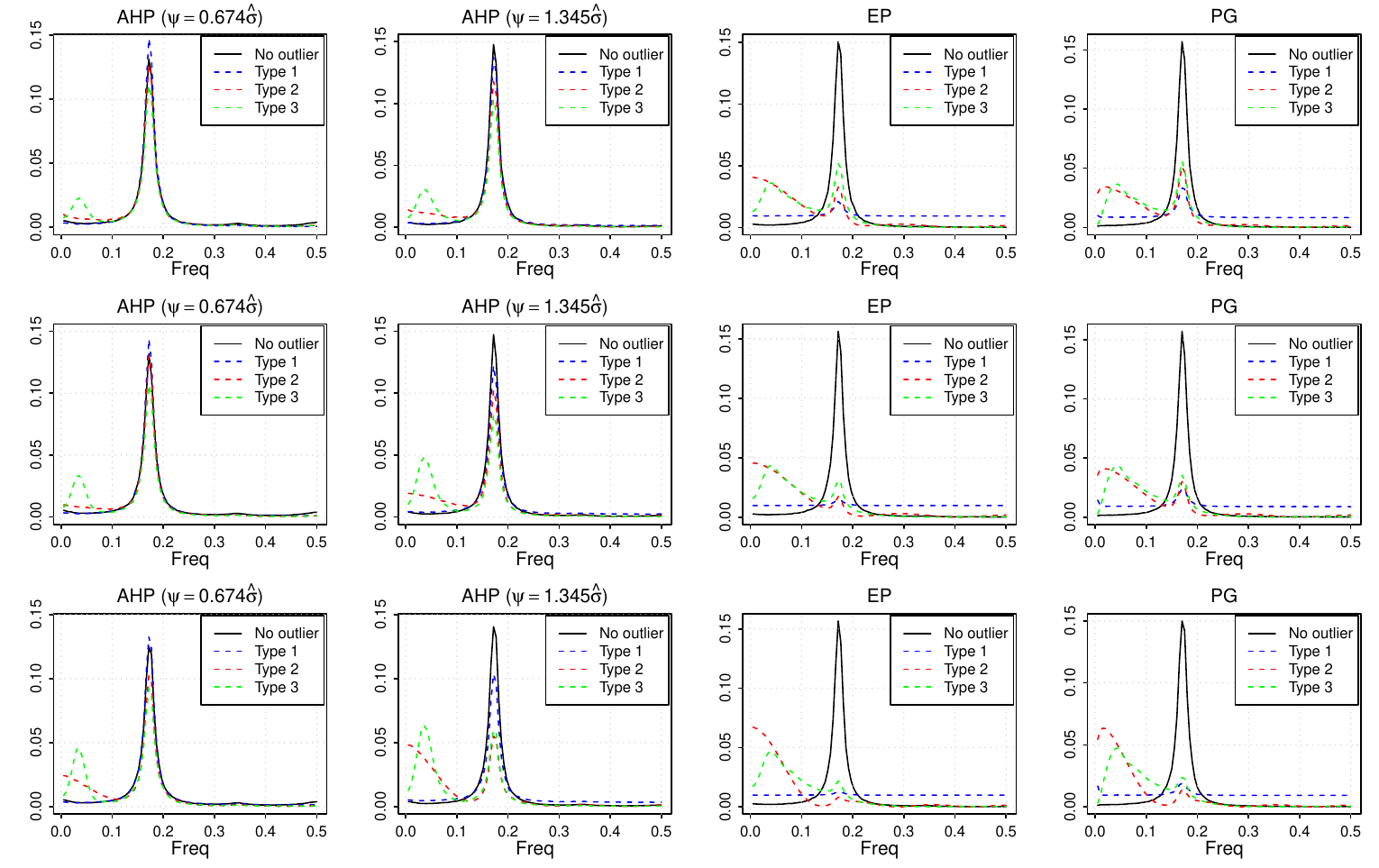}
		\caption{Left to right: the AHPs ($\alpha=0.6, \psi = 0.674\hat{\sigma}$), the AHPs ($\alpha=0.6, \psi = 1.345\hat{\sigma}$), the EP ($\alpha=0.6$), and the PG, respectively. Top to bottom: $(c_1,c_2,c_3)$ is set to $(20, 10, 10)$, $(30, 15, 15)$ and $(40, 20, 20)$, respectively. The periodograms are averaged over 1{,}000 realizations with $n=200$.}
		\label{fisher}
	\end{figure}

	We observe that periodicity is easily detected by all periodogram types in the absence of outliers, with all values in the top two rows of Tables \ref{tb1}--\ref{tb3} close to 1.  By contrast, when outliers are introduced, the PD of the EP and PG decreases rapidly, yielding large differences, while the PD of the AHP remains comparatively stable, leading to smaller discrepancies. The small negative differences (e.g., $-0.002$ for the AHP under Type~1 contamination in Table~\ref{tb1}) are attributable to Monte Carlo error; with 1,000 replications, the standard error of the estimated PD is at most $\sqrt{0.5\times0.5/1000}\approx 0.016$, so values within $\pm 0.03$ of the uncontaminated PD are within sampling variability. Fig.~\ref{fisher} depicts the averaged periodograms of 1{,}000 simulations, illustrating that the dominant peak gradually vanishes in the EP and PG as the outlier magnitude increases, but remains large in the AHP. Notably, for model (\ref{type1}), both the EP and PG become nearly constant, losing all spectral features. We also observe that, as $\psi$ increases (the EP can be viewed as a limiting case of the AHP with $\psi \to \infty$), the AHP loses robustness but gains computational efficiency. Specifically, the average computation times for different values of $\psi$ are reported at the bottom of Table \ref{tb1}.  All computations were performed on a PC equipped with an Intel Core i9-13900KF CPU @5.2 GHz on a single thread and 64 GB of RAM. 
	
	\begin{table}[htb]
		\centering
		\spacingset{1}
		\caption{Fisher's test under Type 1 contamination.}
		\label{tb1}
		\footnotesize
		\setlength{\tabcolsep}{5pt}
		\renewcommand{\arraystretch}{1}
		\begin{tabular}{lllcccccc}
			\toprule[1.2pt]
			& & \shortstack{Significance\\level} & $\alpha$ & \multicolumn{3}{c}{AHP} & EP & PG \\
			& & & & $\psi=0.674\hat\sigma$ & $\psi=0.935\hat\sigma$ & $\psi=1.345\hat\sigma$ & & \\
			\midrule
			\multirow{4}{*}{$y_t$} & \multirow{4}{*}{No outliers} & 0.01 & \multirow{2}{*}{0.6} & 0.997 & 0.998 & 1.000 & 1.000 & 1.000 \\
			& & 0.05 & & 1.000 & 1.000 & 1.000 & 1.000 & 1.000 \\
			& & 0.01 & \multirow{2}{*}{0.8} & 0.977 & 0.988 & 0.995 & 1.000 & 1.000 \\
			& & 0.05 & & 0.998 & 1.000 & 1.000 & 1.000 & 1.000 \\
			\midrule
			\multirow{12}{*}{$y_{1,t}$} & \multirow{4}{*}{$c_1=20\hat\sigma$} & 0.01 & \multirow{2}{*}{0.6} & 0.999 & 1.000 & 1.000 & 0.380 & 0.641 \\
			& & 0.05 & & 1.000 & 1.000 & 1.000 & 0.570 & 0.809 \\
			& & 0.01 & \multirow{2}{*}{0.8} & 0.989 & 0.994 & 0.994 & 0.015 & 0.641 \\
			& & 0.05 & & 0.999 & 1.000 & 1.000 & 0.049 & 0.809 \\
			& \multirow{4}{*}{$c_1=30\hat\sigma$} & 0.01 & \multirow{2}{*}{0.6} & 1.000 & 1.000 & 1.000 & 0.055 & 0.198 \\
			& & 0.05 & & 1.000 & 1.000 & 1.000 & 0.148 & 0.359 \\
			& & 0.01 & \multirow{2}{*}{0.8} & 0.994 & 0.994 & 0.991 & 0.000 & 0.198 \\
			& & 0.05 & & 1.000 & 1.000 & 0.999 & 0.000 & 0.359 \\
			& \multirow{4}{*}{$c_1=40\hat\sigma$} & 0.01 & \multirow{2}{*}{0.6} & 1.000 & 1.000 & 0.999 & 0.004 & 0.037 \\
			& & 0.05 & & 1.000 & 1.000 & 1.000 & 0.018 & 0.106 \\
			& & 0.01 & \multirow{2}{*}{0.8} & 0.994 & 0.993 & 0.982 & 0.000 & 0.037 \\
			& & 0.05 & & 1.000 & 1.000 & 0.994 & 0.000 & 0.106 \\
			\midrule
			\multirow{12}{*}{Difference} & \multirow{4}{*}{$c_1=20\hat\sigma$} & 0.01 & \multirow{2}{*}{0.6} & $-0.002$ & $-0.002$ & 0.000 & 0.620 & 0.359 \\
			& & 0.05 & & 0.000 & 0.000 & 0.000 & 0.430 & 0.191 \\
			& & 0.01 & \multirow{2}{*}{0.8} & $-0.012$ & $-0.006$ & 0.001 & 0.985 & 0.359 \\
			& & 0.05 & & $-0.001$ & 0.000 & 0.000 & 0.951 & 0.191 \\
			& \multirow{4}{*}{$c_1=30\hat\sigma$} & 0.01 & \multirow{2}{*}{0.6} & $-0.003$ & $-0.002$ & 0.000 & 0.945 & 0.802 \\
			& & 0.05 & & 0.000 & 0.000 & 0.000 & 0.852 & 0.641 \\
			& & 0.01 & \multirow{2}{*}{0.8} & $-0.017$ & $-0.006$ & 0.004 & 1.000 & 0.802 \\
			& & 0.05 & & $-0.002$ & 0.000 & 0.001 & 1.000 & 0.641 \\
			& \multirow{4}{*}{$c_1=40\hat\sigma$} & 0.01 & \multirow{2}{*}{0.6} & $-0.003$ & $-0.002$ & 0.001 & 0.996 & 0.963 \\
			& & 0.05 & & 0.000 & 0.000 & 0.000 & 0.982 & 0.894 \\
			& & 0.01 & \multirow{2}{*}{0.8} & $-0.017$ & $-0.005$ & 0.013 & 1.000 & 0.963 \\
			& & 0.05 & & $-0.002$ & 0.000 & 0.006 & 1.000 & 0.894 \\
			\toprule[1.2pt]
			\multicolumn{4}{l}{Avg.\ time (s)} & 0.297 & 0.227 & 0.223 & 0.189 & -- \\
			\bottomrule[1.2pt]
		\end{tabular}
		\spacingset{1.8}
	\end{table}
	\begin{table}[ht]
		\centering
		\spacingset{1}
		\caption{Fisher's test under Type 2 contamination.}
		\label{tb2}
		\footnotesize
		\setlength{\tabcolsep}{5pt}
		\renewcommand{\arraystretch}{1}
		\begin{tabular}{lllcccccc}
			\toprule[1.2pt]
			& & \shortstack{Significance\\level} & $\alpha$ & \multicolumn{3}{c}{AHP} & EP & PG \\
			& & & & $\psi=0.674\hat\sigma$ & $\psi=0.935\hat\sigma$ & $\psi=1.345\hat\sigma$ & & \\
			\midrule
			\multirow{4}{*}{$y_t$} & \multirow{4}{*}{No outliers} & 0.01 & \multirow{2}{*}{0.6} & 0.997 & 0.998 & 1.000 & 1.000 & 1.000 \\
			& & 0.05 & & 1.000 & 1.000 & 1.000 & 1.000 & 1.000 \\
			& & 0.01 & \multirow{2}{*}{0.8} & 0.977 & 0.988 & 0.995 & 1.000 & 1.000 \\
			& & 0.05 & & 0.998 & 1.000 & 1.000 & 1.000 & 1.000 \\
			\midrule
			\multirow{12}{*}{$y_{2,t}$} & \multirow{4}{*}{$c_2=10\hat\sigma$} & 0.01 & \multirow{2}{*}{0.6} & 0.999 & 0.999 & 0.996 & 0.103 & 0.325 \\
			& & 0.05 & & 1.000 & 1.000 & 1.000 & 0.238 & 0.521 \\
			& & 0.01 & \multirow{2}{*}{0.8} & 0.940 & 0.931 & 0.819 & 0.000 & 0.325 \\
			& & 0.05 & & 0.984 & 0.977 & 0.930 & 0.000 & 0.521 \\
			& \multirow{4}{*}{$c_2=15\hat\sigma$} & 0.01 & \multirow{2}{*}{0.6} & 0.999 & 0.996 & 0.976 & 0.000 & 0.012 \\
			& & 0.05 & & 1.000 & 1.000 & 0.997 & 0.001 & 0.044 \\
			& & 0.01 & \multirow{2}{*}{0.8} & 0.932 & 0.836 & 0.491 & 0.000 & 0.012 \\
			& & 0.05 & & 0.979 & 0.942 & 0.709 & 0.000 & 0.044 \\
			& \multirow{4}{*}{$c_2=20\hat\sigma$} & 0.01 & \multirow{2}{*}{0.6} & 0.998 & 0.981 & 0.912 & 0.000 & 0.000 \\
			& & 0.05 & & 1.000 & 0.999 & 0.973 & 0.000 & 0.000 \\
			& & 0.01 & \multirow{2}{*}{0.8} & 0.874 & 0.640 & 0.148 & 0.000 & 0.000 \\
			& & 0.05 & & 0.956 & 0.813 & 0.315 & 0.000 & 0.000 \\
			\midrule
			\multirow{12}{*}{Difference} & \multirow{4}{*}{$c_2=10\hat\sigma$} & 0.01 & \multirow{2}{*}{0.6} & $-0.002$ & $-0.001$ & 0.004 & 0.897 & 0.675 \\
			& & 0.05 & & 0.000 & 0.000 & 0.000 & 0.762 & 0.479 \\
			& & 0.01 & \multirow{2}{*}{0.8} & 0.037 & 0.057 & 0.176 & 1.000 & 0.675 \\
			& & 0.05 & & 0.014 & 0.023 & 0.070 & 1.000 & 0.479 \\
			& \multirow{4}{*}{$c_2=15\hat\sigma$} & 0.01 & \multirow{2}{*}{0.6} & $-0.002$ & 0.002 & 0.024 & 1.000 & 0.988 \\
			& & 0.05 & & 0.000 & 0.000 & 0.000 & 0.992 & 0.956 \\
			& & 0.01 & \multirow{2}{*}{0.8} & 0.045 & 0.152 & 0.504 & 1.000 & 0.988 \\
			& & 0.05 & & 0.019 & 0.032 & 0.201 & 1.000 & 0.956 \\
			& \multirow{4}{*}{$c_2=20\hat\sigma$} & 0.01 & \multirow{2}{*}{0.6} & $-0.001$ & 0.017 & 0.088 & 1.000 & 1.000 \\
			& & 0.05 & & 0.000 & 0.001 & 0.027 & 1.000 & 1.000 \\
			& & 0.01 & \multirow{2}{*}{0.8} & 0.103 & 0.348 & 0.847 & 1.000 & 1.000 \\
			& & 0.05 & & 0.042 & 0.187 & 0.685 & 1.000 & 1.000 \\
			\bottomrule[1.2pt]
		\end{tabular}
		\spacingset{1.8}
	\end{table}
	
	\begin{table}[ht]
		\centering
		\spacingset{1}
		\caption{Fisher's test under Type 3 contamination.}
		\label{tb3}
		\footnotesize
		\setlength{\tabcolsep}{5pt}
		\renewcommand{\arraystretch}{1}
		\begin{tabular}{lllcccccc}
			\toprule[1.2pt]
			& & \shortstack{Significance\\level} & $\alpha$ & \multicolumn{3}{c}{AHP} & EP & PG \\
			& & & & $\psi=0.674\hat\sigma$ & $\psi=0.935\hat\sigma$ & $\psi=1.345\hat\sigma$ & & \\
			\midrule
			\multirow{4}{*}{$y_t$} & \multirow{4}{*}{No outliers} & 0.01 & \multirow{2}{*}{0.6} & 0.997 & 0.998 & 1.000 & 1.000 & 1.000 \\
			& & 0.05 & & 1.000 & 1.000 & 1.000 & 1.000 & 1.000 \\
			& & 0.01 & \multirow{2}{*}{0.8} & 0.977 & 0.988 & 0.995 & 1.000 & 1.000 \\
			& & 0.05 & & 0.998 & 1.000 & 1.000 & 1.000 & 1.000 \\
			\midrule
			\multirow{12}{*}{$y_{3,t}$} & \multirow{4}{*}{$c_3=10\hat\sigma$} & 0.01 & \multirow{2}{*}{0.6} & 0.993 & 0.988 & 0.977 & 0.528 & 0.675 \\
			& & 0.05 & & 1.000 & 1.000 & 0.998 & 0.740 & 0.845 \\
			& & 0.01 & \multirow{2}{*}{0.8} & 0.857 & 0.844 & 0.762 & 0.072 & 0.675 \\
			& & 0.05 & & 0.961 & 0.954 & 0.901 & 0.165 & 0.845 \\
			& \multirow{4}{*}{$c_3=15\hat\sigma$} & 0.01 & \multirow{2}{*}{0.6} & 0.986 & 0.972 & 0.899 & 0.100 & 0.193 \\
			& & 0.05 & & 1.000 & 0.994 & 0.974 & 0.214 & 0.348 \\
			& & 0.01 & \multirow{2}{*}{0.8} & 0.790 & 0.674 & 0.360 & 0.000 & 0.193 \\
			& & 0.05 & & 0.934 & 0.886 & 0.686 & 0.001 & 0.348 \\
			& \multirow{4}{*}{$c_3=20\hat\sigma$} & 0.01 & \multirow{2}{*}{0.6} & 0.970 & 0.907 & 0.700 & 0.006 & 0.033 \\
			& & 0.05 & & 0.991 & 0.983 & 0.905 & 0.036 & 0.094 \\
			& & 0.01 & \multirow{2}{*}{0.8} & 0.653 & 0.425 & 0.110 & 0.000 & 0.033 \\
			& & 0.05 & & 0.927 & 0.859 & 0.656 & 0.000 & 0.094 \\
			\midrule
			\multirow{12}{*}{Difference} & \multirow{4}{*}{$c_3=10\hat\sigma$} & 0.01 & \multirow{2}{*}{0.6} & 0.004 & 0.010 & 0.023 & 0.472 & 0.325 \\
			& & 0.05 & & 0.000 & 0.000 & 0.002 & 0.260 & 0.155 \\
			& & 0.01 & \multirow{2}{*}{0.8} & 0.120 & 0.144 & 0.233 & 0.928 & 0.325 \\
			& & 0.05 & & 0.037 & 0.046 & 0.099 & 0.835 & 0.155 \\
			& \multirow{4}{*}{$c_3=15\hat\sigma$} & 0.01 & \multirow{2}{*}{0.6} & 0.011 & 0.026 & 0.101 & 0.900 & 0.807 \\
			& & 0.05 & & 0.000 & 0.006 & 0.026 & 0.786 & 0.652 \\
			& & 0.01 & \multirow{2}{*}{0.8} & 0.187 & 0.314 & 0.635 & 1.000 & 0.807 \\
			& & 0.05 & & 0.064 & 0.114 & 0.314 & 0.999 & 0.652 \\
			& \multirow{4}{*}{$c_3=20\hat\sigma$} & 0.01 & \multirow{2}{*}{0.6} & 0.027 & 0.091 & 0.300 & 0.994 & 0.967 \\
			& & 0.05 & & 0.009 & 0.017 & 0.095 & 0.964 & 0.906 \\
			& & 0.01 & \multirow{2}{*}{0.8} & 0.324 & 0.563 & 0.885 & 1.000 & 0.967 \\
			& & 0.05 & & 0.071 & 0.141 & 0.344 & 1.000 & 0.906 \\
			\bottomrule[1.2pt]
		\end{tabular}
		\spacingset{1.8}
	\end{table}

	\subsection{Coverage Probability of Confidence Intervals}
	\label{sec:coverage}
	
	We evaluate the finite-sample coverage of the pointwise confidence intervals developed in Section~3.2. We generate $n=200$, $500$, and $1000$ observations from the stationary Gaussian AR(2) process
	\[
	y_t=0.9y_{t-1}-0.6y_{t-2}+\epsilon_t,
	\qquad
	\epsilon_t\stackrel{\mathrm{i.i.d.}}{\sim}N(0,1).
	\]
	A sufficiently long burn-in period is discarded before retaining the
	$n$ observations. The marginal variance of this process is
	$\sigma_y^2=16/7$. To match the fixed-threshold framework of
	Theorem~1, we set
	\[
	\psi=0.674\sigma_y
	\qquad\text{and}\qquad
	\psi=1.345\sigma_y,
	\]
	rather than estimating $\psi$ separately in each replication.
	
	For each fixed $(\alpha,\psi)$, the benchmark value
	$g(\lambda,\alpha,\psi)$ at
	$\lambda=0.25\times2\pi$ is obtained by high-precision numerical
	evaluation of $\mu(\alpha,\psi)$, $\eta(\alpha,\psi)$, and the
	autocovariances of the score process
	\[
	V_t=
	\dot\rho_{\alpha,\psi}
	\{y_t-\mu(\alpha,\psi)\}.
	\]
	Specifically, the Fourier series
	\[
	g(\lambda,\alpha,\psi)
	=
	\eta^2(\alpha,\psi)
	\left[
	\gamma(0,\alpha,\psi)
	+
	2\sum_{\tau=1}^{\infty}
	\gamma(\tau,\alpha,\psi)\cos(\lambda\tau)
	\right]
	\]
	is evaluated numerically using a sufficiently large lag truncation.
	
	For each of 1,000 independent replications, we construct a 95\% pointwise interval based on the unnormalized raw AHP $\mathrm{CI}_1$ and approximate intervals based on the unnormalized smoothed AHP $\mathrm{CI}_2$, using equal rectangular weights with $M_n=3$, $5$, and $11$. Table~\ref{tb:coverage} reports the empirical coverage probabilities for $\alpha=0.6$ and $0.8$. The Monte Carlo standard error of an estimated coverage probability near $0.95$ is approximately $0.007$.
	
	The raw-AHP interval exhibits moderate undercoverage at $n=200$, while its empirical coverage moves closer to the nominal level as the sample size increases, which is consistent with the asymptotic nature of the chi-square approximation. The approximate smoothed-AHP intervals with larger smoothing spans exhibit conservative coverage in the considered settings. Since these intervals rely on the
	local-independence and common-local-scale approximation introduced in Section \ref{CIs}, rather than on a formal limiting distribution for the smoothed AHP, no specific theoretical explanation for the observed conservativeness is asserted. The different values of $M_n$ are included as a finite-sample sensitivity analysis; theoretically optimal bandwidth selection is beyond the scope of the present paper.
	\begin{table}[ht]
		\centering
		\spacingset{1}
		\caption{Empirical coverage probabilities (95\% nominal level) of CIs for the AHS at
			$0.25\times2\pi$. Results are based on 1,000 replications of a Gaussian
			AR(2) process.}
		\label{tb:coverage}
		\footnotesize
		\setlength{\tabcolsep}{4pt}
		\renewcommand{\arraystretch}{1.15}
		\begin{tabular}{ccccccccc}
			\toprule[1.2pt]
			\multicolumn{2}{c}{}
			& \multicolumn{4}{c}{$\psi=0.674\hat{\sigma}$}
			& \multicolumn{3}{c}{$\psi=1.345\hat{\sigma}$} \\
			\cmidrule(lr){3-6}
			\cmidrule(lr){7-9}
			$\alpha$
			& $n$
			& CI$_1$
			& \shortstack{CI$_2$\\ ($M_n=3$)}
			& \shortstack{CI$_2$\\ ($M_n=5$)}
			& \shortstack{CI$_2$\\ ($M_n=11$)}
			& CI$_1$
			& \shortstack{CI$_2$\\ ($M_n=5$)}
			& \shortstack{CI$_2$\\ ($M_n=11$)} \\
			\midrule
			\multirow{3}{*}{$0.6$}
			& 200  & 0.931 & 0.938 & 0.952 & 0.971 & 0.934 & 0.949 & 0.969 \\
			& 500  & 0.942 & 0.947 & 0.956 & 0.978 & 0.943 & 0.953 & 0.975 \\
			& 1000 & 0.948 & 0.951 & 0.957 & 0.981 & 0.947 & 0.954 & 0.979 \\
			\midrule
			\multirow{3}{*}{$0.8$}
			& 200  & 0.929 & 0.935 & 0.950 & 0.968 & 0.931 & 0.947 & 0.966 \\
			& 500  & 0.941 & 0.945 & 0.954 & 0.976 & 0.942 & 0.951 & 0.973 \\
			& 1000 & 0.947 & 0.949 & 0.956 & 0.980 & 0.946 & 0.953 & 0.977 \\
			\bottomrule[1.2pt]
		\end{tabular}
		\spacingset{1.8}
	\end{table}
	
	\section{Applications}\label{APP}
	We consider two applications to demonstrate the AHP's ability to reveal hidden periodicities and cluster time series. %and retain spectral structure in the presence of outliers.
	
	\subsection{Detection of Hidden Periodicities}
\begin{figure}[!htbp]
		\centering
		
		\includegraphics[width=\textwidth]{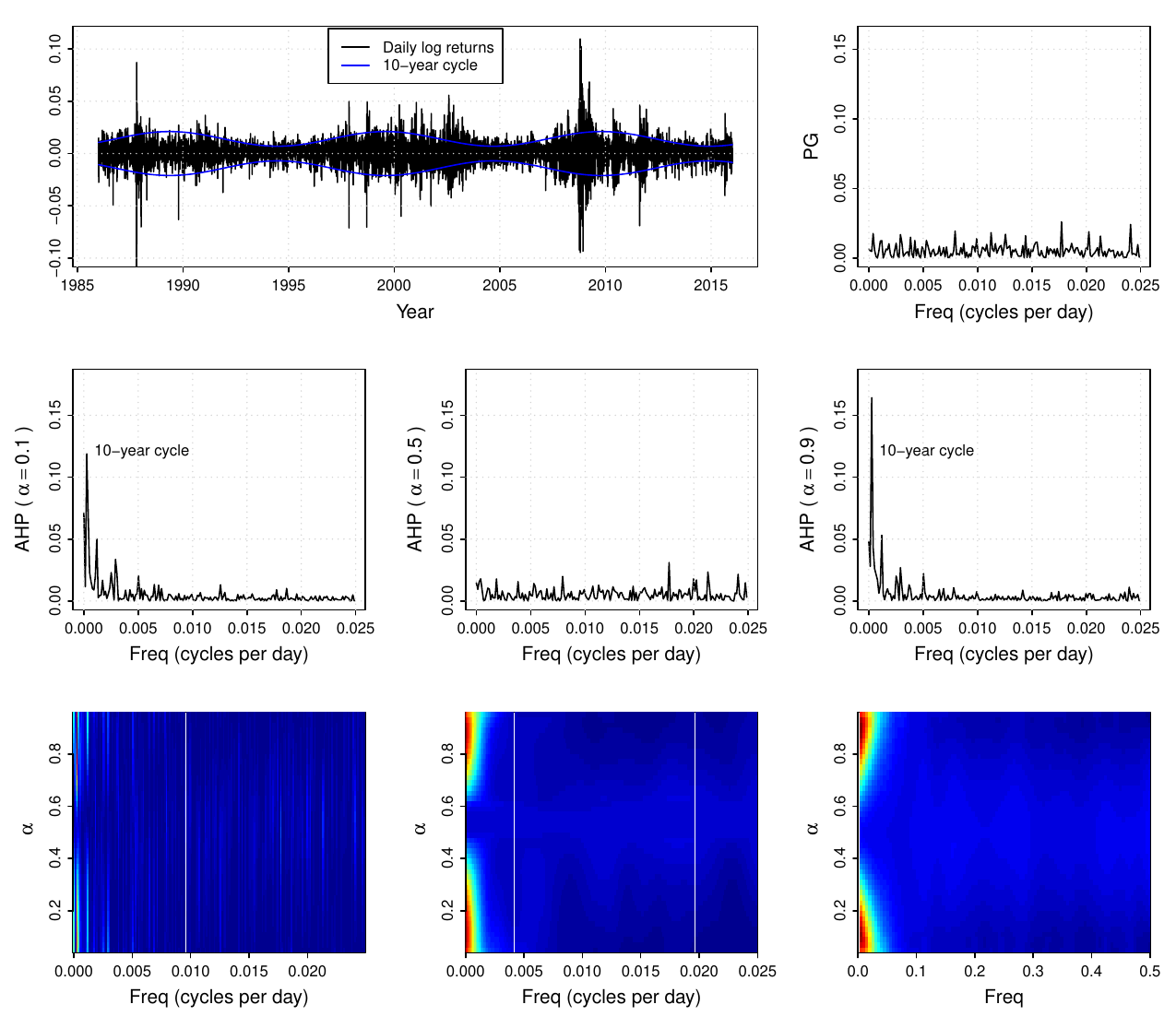}
		\caption{Top row (left): daily log returns of the S$\&$P 500 Index, along with its 10-year cycle; top row (right): the PG. Second row: the AHPs at $\alpha=0.1, 0.5$ and $0.9$, respectively. Bottom row (left): the AHP at $\alpha=0.05, 0.07,\cdots, 0.95$; bottom row (middle): the AHS estimator obtained by smoothing the AHP; bottom row (right): the Monte Carlo estimate of the AHS for model (\ref{gar}). The AHPs are computed at $\psi = 1.345\hat{\sigma}$.}
		\label{sp}
	\end{figure}

\begin{figure}[!htbp]
		\centering
		
		\includegraphics[width=0.8\textwidth]{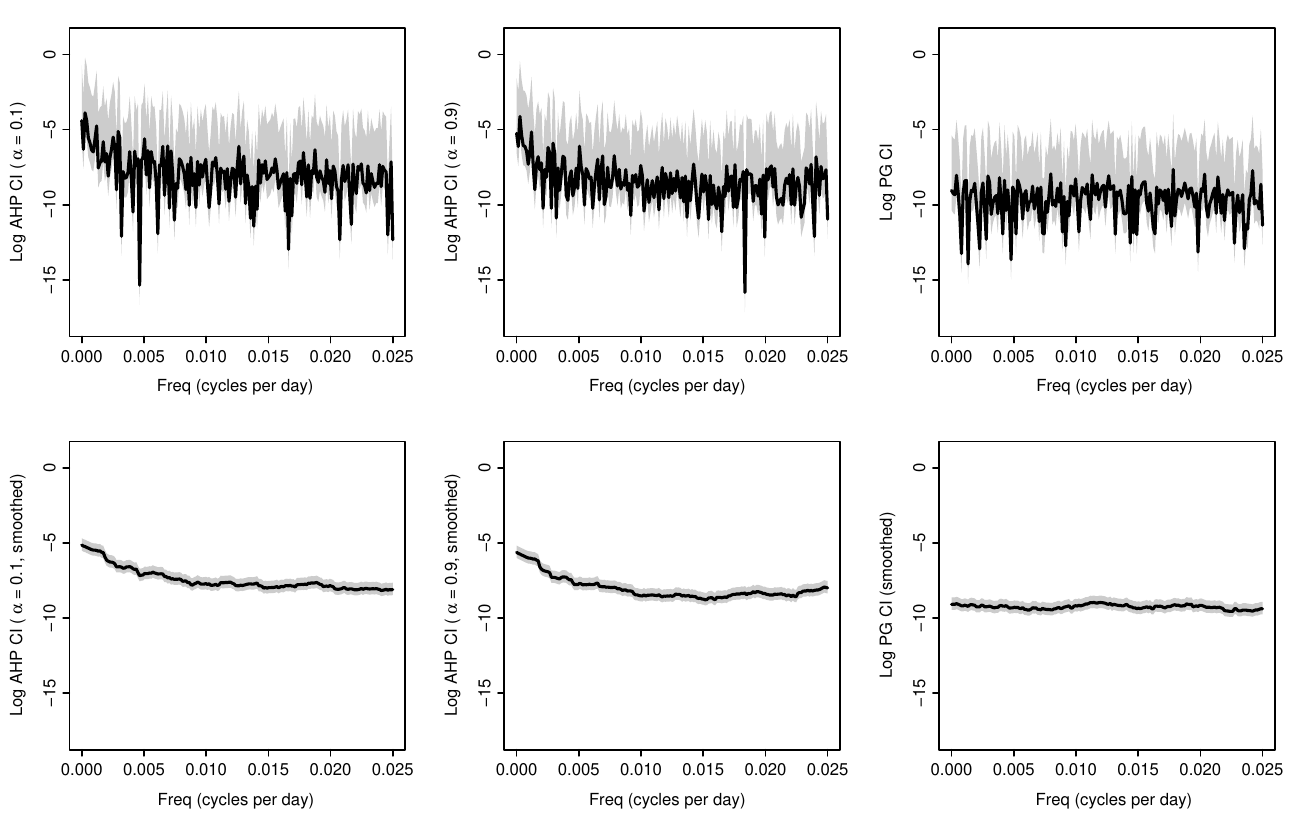}
		\caption{Top row: log spectral estimates and pointwise confidence
			intervals based on the raw periodograms. Bottom row: the corresponding
			results based on the smoothed periodograms. The black curves denote
			the log spectral estimates, and the gray bands denote the 95\%
			pointwise confidence intervals. Equal smoothing weights are used with
			$M_n=11$ and $L_n=23$.}
		\label{log}
	\end{figure}

	%In the first example, 
	First, we examine the daily log returns of the S\&P 500 Index over the period 1986–2015.  The data are publicly available at \url{finance.yahoo.com} or \url{fred.stlouisfed.org}. As shown in Fig.~\ref{sp}, the AHPs at $\alpha = 0.1$ and $0.9$ exhibit a pronounced low-frequency feature corresponding approximately to a 10-year time scale. In contrast, the PG produces a nearly featureless, flat spectrum, similar to that produced by the AHP at $\alpha = 0.5$.  We further compute the smoothed AHP at $\alpha=0.05, 0.07,\cdots, 0.95$ and observe that the spectral feature is similar to the AHS of a GARCH(1,1) model
	\citep{engle1982autoregressive,bollerslev1986generalized}:
	\begin{equation}\label{gar}
		y_t=\sigma_t\varepsilon_t,
		\qquad
		\varepsilon_t \overset{\mathrm{i.i.d.}}{\sim} N(0,1),
	\end{equation}
	where $\sigma_t^2=10^{-6}+0.49y^2_{t-1}+0.49\sigma_{t-1}^2$. The Monte Carlo estimate of the AHS exhibits the spectral patterns of the GARCH(1,1) model at both lower and upper $\alpha$ levels in the low-frequency range, whereas its ordinary spectrum is a constant. The AHS is estimated by averaging $5{,}000$ smoothed AHPs. In Fig.~\ref{log}, we further show the logarithmic CIs defined in \eqref{log1} and \eqref{log2}. As can be seen, smoothing yields a more stable estimate of the spectrum, and the resulting intervals support the presence of a low-frequency feature corresponding to approximately a 10-year time scale. For Fig.~\ref{log}, the confidence intervals are constructed from the unnormalized AHPs. Normalization introduces a random denominator, which changes the chi-square distribution and induces dependence among the ordinates. Therefore, the confidence intervals in Section \ref{CIs} cannot be applied directly to the normalized AHPs.
	
	
	\subsection{Time Series Clustering}
	%In the second example, 
	Second, we cluster the return series of S\&P 500 constituents. We focus on companies in the Financials and Utilities sectors, which exhibit distinct economic characteristics and therefore provide a meaningful benchmark for evaluating the AHP's ability to identify sector-specific spectral structures. The sample period spans from February 8, 2013 to February 7, 2018. To ensure comparability across companies, we exclude firms with incomplete price records over the sample period. The resulting dataset contains 77 companies, including 51 companies from the Financials sector and 26 companies from the Utilities sector.\footnote{The dataset is available at \url{https://www.kaggle.com/camnugent/sandp500}.}
	
	Each company is represented by a univariate time series of daily closing prices. We transform the original price series into daily log returns. After transformation, all return series have a common length of 1258. Finally, each return series is standardized to have zero mean and unit variance. Besides the AHP ($\psi = 1.345\hat{\sigma}, \quad \alpha = 0.1,0.3,0.5,0.7,0.9$), we also consider three competitive spectral estimators: the PG, the QP (computed at quantiles $0.1,0.3,0.5,0.7,0.9$), and the Wavelet features (WLT) \citep{percival2000wavelet,d2014wavelet}. For WLT, we apply the MODWT using the LA(8) wavelet filter and construct a feature vector. Then, we compute the distance matrices for the four methods using RMSE. The resulting distance matrices are then used as input to hierarchical clustering using Ward's linkage (\texttt{method = \textquotesingle ward.D2\textquotesingle}), and each dendrogram is cut into two clusters. The clustering results are evaluated against the reference sector labels using the adjusted Rand index (ARI) \citep{vinh2009information} and the similarity index (SIM) implemented in the \texttt{R} package \texttt{TSClust} \citep{montero2015tsclust}.
	
	Table \ref{clu1} summarizes the clustering performance of the four methods. The AHP achieves the best overall performance, with a similarity index of 0.955 and an ARI of 0.846. The PG and QP also perform well, attaining ARI values of 0.749 and 0.703, respectively, although both are outperformed by the AHP. By contrast, the WLT yields substantially lower values for both measures, with a similarity index of 0.439 and an ARI of 0.025, indicating that its clustering result agrees only weakly with the reference sector labels.
	
	Table \ref{clu2} provides a more detailed view of the cluster assignments. Under the AHP-based clustering, all 51 Financials companies are assigned to Cluster 1, together with only 3 Utilities companies, whereas Cluster 2 consists exclusively of the remaining 23 Utilities companies. The PG and QP produce similar but slightly less accurate partitions: their Financials-dominated clusters contain 5 and 6 Utilities companies, respectively, while their second clusters consist entirely of Utilities companies. In contrast, the WLT assigns all but one company to Cluster 1, resulting in a highly unbalanced partition that fails to effectively separate the two sectors. Overall, these results indicate that the AHP captures sector-specific spectral characteristics more effectively than the competing methods, while the PG and QP also retain substantial sector-discriminative information.
	
\begin{figure}[htbp]
		\centering
		\includegraphics[width=0.9\textwidth]{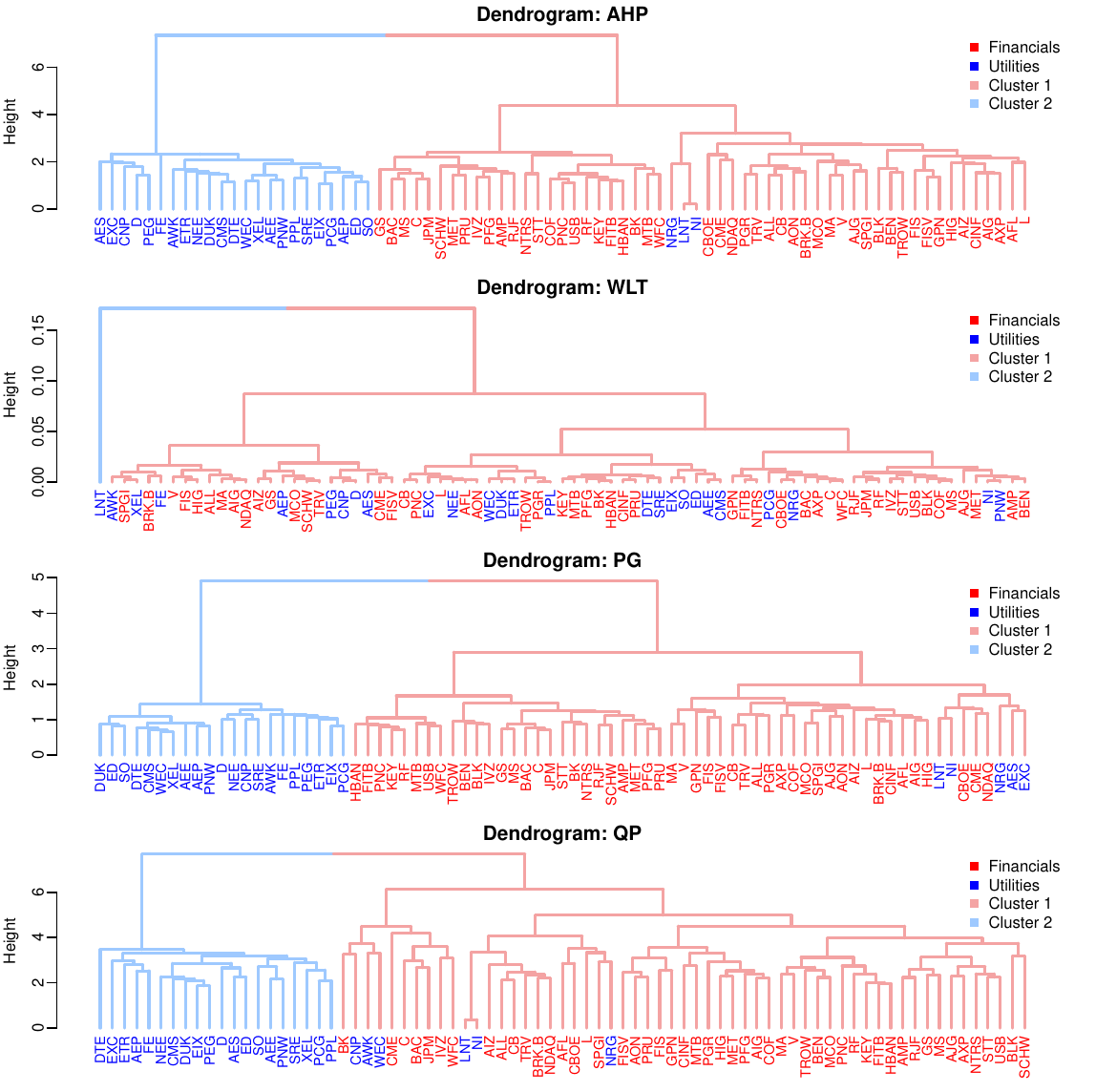}
		\caption{Dendrograms obtained from Ward's hierarchical clustering
			based on the AHP, WLT, PG, and QP dissimilarities. Leaf-label colors
			indicate the reference sectors, whereas branch colors indicate the
			two clusters obtained by cutting each dendrogram at $K=2$. Cluster
			labels are method-specific.}
		\label{dend1}
	\end{figure}

	Finally, Fig.~\ref{dend1} presents the dendrograms obtained from the four methods. The AHP produces the clearest sector-aligned hierarchical structure. Most Utilities companies form a compact branch that remains separate from the Financials-dominated branch until a relatively high level of merging, while only a small number of Utilities companies are embedded in the latter branch. The PG and QP exhibit broadly similar patterns, but the separation between the two sectors is less pronounced, with more Utilities companies grouped together with Financials companies. In contrast, the WLT produces a highly unbalanced hierarchy in which nearly all companies are grouped into a single major cluster, providing little evidence of sector-specific separation. These dendrograms are consistent with the quantitative results in Tables \ref{clu1} and \ref{clu2}, further indicating that the AHP most effectively captures the sector-related spectral structure.

	\begin{table}[htbp]
		\centering
		\caption{Clustering results.}
		\label{clu1}
		\begin{tabular}{ccccc}
			\toprule[1.2pt]
			& AHP & WLT & PG & QP  \\
			\midrule
			SIM  & \textbf{0.955}& 0.439 & 0.923  & 0.907   \\
			ARI & \textbf{0.846} & 0.025 & 0.749  &0.703   \\
			\bottomrule[1.2pt]
		\end{tabular}
	\end{table}
	
	\begin{table}[htbp]
		\centering
		\caption{Number of companies from each sector assigned to each cluster (cluster labels are method-specific and are not
			directly comparable across methods).}
		\label{clu2}
		
		% 第一行
		\begin{subtable}[t]{0.4\textwidth}
			\centering
			\caption{AHP}
			\begin{tabular}{lcc}
				\toprule
				& Financials & Utilities \\
				\midrule
				Cluster 1 & 51 & 3  \\
				Cluster 2 & 0  & 23 \\
				\bottomrule
			\end{tabular}
		\end{subtable}
		\hfill
		\begin{subtable}[t]{0.4\textwidth}
			\centering
			\caption{WLT}
			\begin{tabular}{lcc}
				\toprule
				& Financials & Utilities \\
				\midrule
				Cluster 1 & 51 & 25  \\
				Cluster 2 & 0 & 1 \\
				\bottomrule
			\end{tabular}
		\end{subtable}
		
		\par\vspace{0.5cm}
		
		% 第二行
		\begin{subtable}[t]{0.4\textwidth}
			\centering
			\caption{PG}
			\begin{tabular}{lcc}
				\toprule
				& Financials & Utilities \\
				\midrule
				Cluster 1 & 51 & 5  \\
				Cluster 2 & 0  & 21 \\
				\bottomrule
			\end{tabular}
		\end{subtable}
		\hfill
		\begin{subtable}[t]{0.4\textwidth}
			\centering
			\caption{QP}
			\begin{tabular}{lcc}
				\toprule
				& Financials & Utilities \\
				\midrule
				Cluster 1 & 51 & 6 \\
				Cluster 2 & 0 & 20  \\
				\bottomrule
			\end{tabular}
		\end{subtable}
	\end{table}

	\section{Conclusion}\label{CO}
	We have proposed the AHP as a nonparametric spectral M-estimator for time series analysis. The AHP generalizes the PG, LP, QP, EP, and HP through the choice of two parameters: the asymmetry parameter $\alpha\in(0,1)$ and the threshold parameter $\psi>0$ (see Table~\ref{tab:comparison} for a systematic summary). We established the asymptotic theory showing that the raw AHP ordinates are asymptotically distributed as scaled chi-square random variables, and that a suitably smoothed AHP provides a consistent estimator for the AHS. The AHS itself is defined as a scaled version of the ordinary power spectrum of the asymmetrically truncated and weighted residual process $\{\dot\rho_{\alpha,\psi}(y_t-\mu(\alpha,\psi))\}$, providing a clear interpretation in terms of frequency-domain serial dependence. Building on the asymptotic distribution, we developed pointwise confidence intervals and a Fisher-type test for detecting dominant periodic components. Simulations and two real-data applications demonstrate the AHP's capability to detect hidden periodicities and its robustness to outliers, while revealing distributional features that are invisible to the PG.

	\setlength{\bibsep}{0.25em} 
	\spacingset{0.5}
	\bibliographystyle{chicago}
	\bibliography{reff_cleaned}
	
	\spacingset{1.8}
	
	\clearpage
	
	\section*{Appendix A}
	\subsection*{A.1 Lemma and Proof}
	Let the following assumptions be satisfied:
	\begin{itemize}
		\item[A1.]  Let
		$
		U_t=y_t-\mu(\alpha,\psi).
		$
		There exists some $\epsilon>0$ such that $U_t$ has a marginal density
		$f_U$ satisfying
		$
		\sup_{u\in\mathcal N_\epsilon}f_U(u)<\infty,
		\quad
		\mathcal N_\epsilon
		=
		\bigcup_{a\in\{-\psi,0,\psi\}}
		(a-\epsilon,a+\epsilon).
		$
		In addition,
		$
		\mathrm{E}\left\{
		\ddot\rho_{\alpha,\psi}(U_t)
		\right\}>0.
		$
		
		\item[A2.] The deterministic regressor sequence is uniformly bounded:
		$
		\sup_{n,t}\|\mathbf x_{nt}\|
		\leq C_x<\infty.
		$
		
		\item[A3.]The process $\{y_{t}\}$ has the strong mixing property with mixing numbers $a_{\tau}$ ($\tau=1,2,\ldots$) satisfying $a_{\tau}\to 0$ as $\tau\to\infty$.
		
		\item[A4.] There exists a positive definite matrix \(\mathbf{D}_j\) for all $j$ such that \(\mathbf{D}_{jn}\to\mathbf{D}_j\) as \(n\to\infty\), where 
		$
		\mathbf{D}_{jn}:=n^{-1}\sum_{t=1}^{n}\mathbf{x}_{jnt}\mathbf{x}_{jnt}^{\top}.
		$

		\item[A5.]  There exists  \(\mathbf{V}_{jl}(\alpha,\psi)\) for all $j,l$  such that \(\mathbf{V}_{jln}(\alpha,\psi)\to\mathbf{V}_{jl}(\alpha,\psi)\) as \(n\to\infty\), where 
		\begin{equation}\label{vnj}
			\mathbf{V}_{jln}(\alpha,\psi):=n^{-1}\sum_{t=1}^{n}\sum_{s=1}^{n}\text{Cov}\{\dot{\rho}_{\alpha,\psi}(y_{t}-\mu(\alpha,\psi)),\dot{\rho}_{\alpha,\psi}(y_{s}-\mu(\alpha,\psi))\}\,\mathbf{x}_{jnt}\mathbf{x}_{lns}^{\top},\text{ }\text{ }j,l=1,…,q.
		\end{equation}
		
		\item[A6.] A central limit theorem is valid for \(\text{vec}[\boldsymbol{\zeta}_{jn}(\alpha,\psi)]_{j=1}^q\), i.e., \(\text{vec}[\boldsymbol{\zeta}_{jn}(\alpha,\psi)]_{j=1}^q\xrightarrow{D}N(\mathbf{0},[\mathbf{V}_{jl}(\alpha,\psi)]_{j,l=1}^q)\) as \(n\to\infty\).

	\end{itemize}
	
	\begin{lemma} %(Asymmetric Huber Regression). 
		If (A1)--(A6) are satisfied, then
		\[
		\sqrt{n}\text{vec}[\hat{\boldsymbol{\beta}}_{jn}(\omega_{\nu_j},\alpha,\psi)-\boldsymbol{\beta}_{0}(\alpha,\psi)]_{j=1}^q \xrightarrow{D}N(\mathbf{0},[\mathbf{\Lambda}^{-1}_{j}(\alpha,\psi)\mathbf{V}_{jl}(\alpha,\psi)\mathbf{\Lambda}_{l}^{-1}(\alpha,\psi)]_{j,l=1}^q),
		\]
		where  
		$
		\hat{\bm{\beta}}_{jn}(\omega_{\nu_j},\alpha,\psi) := \arg\min_{\bm{\beta}_j \in \mathbb{R}^p} \sum_{t=1}^{n} \rho_{\alpha,\psi}(y_t - \mathbf{x}_{jnt}^T \bm{\beta}_j),
		$
		$\boldsymbol{\beta}_{0}(\alpha,\psi):=[\mu(\alpha,\psi),0,\ldots,0]^{\top}\in\mathbb{R}^{p}$, and 
		\[
		\mathbf{\Lambda}_j(\alpha,\psi):=\lim_{n\to\infty}\mathbf{\Lambda}_{jn}(\alpha,\psi):=\lim_{n\to\infty} \eta^{-1}(\alpha,\psi)\, \mathbf{D}_{jn}   =\eta^{-1}(\alpha,\psi)\, \mathbf{D}_j. 
		\]
	\end{lemma}
	\noindent
	
	\begin{proof}
		First, consider the case $q=1$, for which we suppress the frequency
		subscript $j$ for notational simplicity. Let
		$
		U_t
		:=
		y_t-\mu(\alpha,\psi)
		=
		y_t-\mathbf{x}_{nt}^{\top}
		\boldsymbol{\beta}_0(\alpha,\psi),
		$
		and, for a fixed
		$\boldsymbol{\delta}\in\mathbb{R}^{p}$, define
		$
		c_{nt}(\boldsymbol{\delta})
		:=
		n^{-1/2}\mathbf{x}_{nt}^{\top}\boldsymbol{\delta}.
		$
		Consider the random convex function
		\[
		Z_n(\boldsymbol{\delta})
		:=
		\sum_{t=1}^{n}
		\left\{
		\rho_{\alpha,\psi}
		\bigl(U_t-c_{nt}(\boldsymbol{\delta})\bigr)
		-
		\rho_{\alpha,\psi}(U_t)
		\right\}.
		\]
		
		Under the reparameterization
		$
		\boldsymbol{\delta}
		=
		\sqrt{n}
		\left\{
		\boldsymbol{\beta}
		-
		\boldsymbol{\beta}_0(\alpha,\psi)
		\right\},
		$
		we  write
		\[
		Z_n(\boldsymbol{\delta})
		=
		\sum_{t=1}^{n}
		\left[
		\rho_{\alpha,\psi}
		\left(
		y_t-\mathbf{x}_{nt}^{\top}\boldsymbol{\beta}
		\right)
		-
		\rho_{\alpha,\psi}(U_t)
		\right].
		\]
		Therefore, the minimizer of
		$Z_n(\boldsymbol{\delta})$ satisfies
		\[
		\hat{\boldsymbol{\delta}}_n
		:=
		\arg\min_{\boldsymbol{\delta}\in\mathbb{R}^{p}}
		Z_n(\boldsymbol{\delta})
		=
		\sqrt{n}
		\left\{
		\hat{\boldsymbol{\beta}}_n(\omega_{\nu},\alpha,\psi)
		-
		\boldsymbol{\beta}_0(\alpha,\psi)
		\right\}.
		\]
		We aim to prove
		\begin{equation}\label{delta}
			\hat{\boldsymbol{\delta}}_n
			=
			\boldsymbol{\Lambda}_n^{-1}
			\boldsymbol{\zeta}_n
			+
			o_P(1).
		\end{equation}
		Since $(\alpha,\psi)$ is fixed, its dependence is suppressed from
		$\boldsymbol{\Lambda}_n$ and $\boldsymbol{\zeta}_n$ whenever no
		confusion arises.
		
		Define
		$
		\widetilde Z_n(\boldsymbol{\delta})
		:=
		-
		\boldsymbol{\delta}^{\top}\boldsymbol{\zeta}_n
		+
		\frac{1}{2}
		\boldsymbol{\delta}^{\top}
		\boldsymbol{\Lambda}_n
		\boldsymbol{\delta}.
		$
		We first establish that, for every fixed
		$\boldsymbol{\delta}\in\mathbb{R}^{p}$,
		\begin{equation}\label{zn}
			Z_n(\boldsymbol{\delta})
			=
			\widetilde Z_n(\boldsymbol{\delta})
			+
			o_P(1).
		\end{equation}
		For simplicity, write
		$c_{nt}=c_{nt}(\boldsymbol{\delta})$. Define
		$
		\hat{\boldsymbol{\Lambda}}_n
		:=
		n^{-1}
		\sum_{t=1}^{n}
		\ddot\rho_{\alpha,\psi}(U_t)
		\mathbf{x}_{nt}\mathbf{x}_{nt}^{\top}.
		$
		The values of $\ddot\rho_{\alpha,\psi}$ at
		$-\psi$, $0$, and $\psi$ may be defined arbitrarily, since these
		points have probability zero under \textnormal{(A1)}. By the
		definition of $\eta(\alpha,\psi)$,
		\[
		\mathrm{E}\left\{
		\hat{\boldsymbol{\Lambda}}_n
		\right\}
		=
		\mathrm{E}\left\{
		\ddot\rho_{\alpha,\psi}(U_t)
		\right\}
		\left(
		n^{-1}\sum_{t=1}^{n}
		\mathbf{x}_{nt}\mathbf{x}_{nt}^{\top}
		\right)
		=
		\eta^{-1}(\alpha,\psi)\mathbf{D}_n
		=
		\boldsymbol{\Lambda}_n.
		\]
		
		We next show that
		\begin{equation}\label{lambda-sample}
			\hat{\boldsymbol{\Lambda}}_n
			-
			\boldsymbol{\Lambda}_n
			=
			o_P(1).
		\end{equation}
		For any pair of component indices $r,s$, let
		\[
		H_{nt}^{(r,s)}
		=
		\left[
		\ddot\rho_{\alpha,\psi}(U_t)
		-
		\mathrm{E}\{\ddot\rho_{\alpha,\psi}(U_t)\}
		\right]
		x_{nt,r}x_{nt,s}.
		\]
		Because $\ddot\rho_{\alpha,\psi}$ and the regressors are bounded,
		there exists a finite constant $C_H$ such that
		$|H_{nt}^{(r,s)}|\leq C_H$. Moreover, since $H_{nt}^{(r,s)}$ is a
		measurable function of $y_t$, the strong-mixing covariance inequality
		gives
		\[
		\left|
		\operatorname{Cov}
		\left(
		H_{nt}^{(r,s)},
		H_{nu}^{(r,s)}
		\right)
		\right|
		\leq
		4C_H^2a_{|t-u|}.
		\]
		Consequently,
		\[
		\begin{aligned}
			\operatorname{Var}
			\left\{
			n^{-1}\sum_{t=1}^{n}H_{nt}^{(r,s)}
			\right\}
			\leq
			\frac{C}{n}
			+
			\frac{C}{n}
			\sum_{\tau=1}^{n-1}a_\tau
			=
			\frac{C}{n}
			+
			C\left(
			\frac{1}{n}\sum_{\tau=1}^{n-1}a_\tau
			\right)
			\longrightarrow0,
		\end{aligned}
		\]
		because $a_\tau\to0$ implies
		$n^{-1}\sum_{\tau=1}^{n-1}a_\tau\to0$. Since $p$ is fixed,
		\eqref{lambda-sample} follows.
		
		Now define the remainder
		\[
		\begin{aligned}
			R_{nt}
			:={}
			\rho_{\alpha,\psi}(U_t-c_{nt})
			-
			\rho_{\alpha,\psi}(U_t)
			+
			c_{nt}\dot\rho_{\alpha,\psi}(U_t)
			-
			\frac{1}{2}
			c_{nt}^{2}
			\ddot\rho_{\alpha,\psi}(U_t).
		\end{aligned}
		\]
		It follows that
		\[
		\begin{aligned}
			Z_n(\boldsymbol{\delta})
			={}&
			-
			\boldsymbol{\delta}^{\top}
			\boldsymbol{\zeta}_n
			+
			\frac{1}{2}
			\boldsymbol{\delta}^{\top}
			\hat{\boldsymbol{\Lambda}}_n
			\boldsymbol{\delta}
			+
			\sum_{t=1}^{n}R_{nt}\\
			={}&
			-
			\boldsymbol{\delta}^{\top}
			\boldsymbol{\zeta}_n
			+
			\frac{1}{2}
			\boldsymbol{\delta}^{\top}
			\boldsymbol{\Lambda}_n
			\boldsymbol{\delta}
			+
			\frac{1}{2}
			\boldsymbol{\delta}^{\top}
			\left(
			\hat{\boldsymbol{\Lambda}}_n
			-
			\boldsymbol{\Lambda}_n
			\right)
			\boldsymbol{\delta}
			+
			\sum_{t=1}^{n}R_{nt}.
		\end{aligned}
		\]
		In view of \eqref{lambda-sample}, it remains to show that
		\begin{equation}\label{srnt}
			\sum_{t=1}^{n}R_{nt}
			=
			o_P(1).
		\end{equation}
		
		Let
		$
		\mathcal K=\{-\psi,0,\psi\}
		$
		denote the set of kink points of
		$\rho_{\alpha,\psi}$, and define the crossing event
		\[
		\mathcal C_{nt}
		:=
		\left\{
		\text{the interval with endpoints }
		U_t-c_{nt}
		\text{ and }U_t
		\text{ contains a point in }\mathcal K
		\right\}.
		\]
		On each of the four intervals determined by $\mathcal K$, the
		asymmetric Huber loss is exactly linear or quadratic. Hence, if
		$U_t-c_{nt}$ and $U_t$ lie in the same interval, then
		$
		R_{nt}=0.
		$
		If the interval between $U_t-c_{nt}$ and $U_t$ crosses a kink point,
		the finite change in the second derivative implies that there exists
		a constant $C_{\alpha,\psi}<\infty$, depending only on
		$(\alpha,\psi)$, such that
		\begin{equation}\label{remainder-bound}
			|R_{nt}|
			\leq
			C_{\alpha,\psi}
			c_{nt}^{2}
			\mathbf{1}(\mathcal C_{nt}).
		\end{equation}
		
		By \textnormal{(A2)},
		\[
		|c_{nt}|
		=
		n^{-1/2}
		\left|
		\mathbf{x}_{nt}^{\top}\boldsymbol{\delta}
		\right|
		\leq
		b_n
		:=
		n^{-1/2}C_x\|\boldsymbol{\delta}\|,
		\]
		uniformly in $t$. Since $\psi>0$ is fixed and $b_n\to0$, for all
		sufficiently large $n$, the interval between $U_t-c_{nt}$ and $U_t$
		can cross at most one kink point. Moreover,
		$
		\mathcal C_{nt}
		\subseteq
		\bigcup_{a\in\mathcal K}
		\left\{
		|U_t-a|\leq b_n
		\right\}.
		$
		
		By \textnormal{(A1)}, the marginal density $f_U$ of $U_t$ is bounded
		in neighborhoods of $-\psi$, $0$, and $\psi$. Therefore, there exists
		a finite constant $C_f$ such that, for all sufficiently large $n$,
		\[
		\begin{aligned}
			\Pr(\mathcal C_{nt})
			\leq
			\sum_{a\in\mathcal K}
			\Pr\left(
			|U_t-a|\leq b_n
			\right)
			\leq
			\sum_{a\in\mathcal K}
			\int_{a-b_n}^{a+b_n}f_U(u)\,du
			\leq
			6C_fb_n
			=
			O(n^{-1/2}),
		\end{aligned}
		\]
		uniformly in $t$. Combining this result with
		\eqref{remainder-bound} gives
		\[
		\begin{aligned}
			\mathrm{E}|R_{nt}|
			\leq
			C_{\alpha,\psi}
			c_{nt}^{2}
			\Pr(\mathcal C_{nt})
			\leq
			C_{\alpha,\psi}
			b_n^{2}
			\Pr(\mathcal C_{nt})
			=
			O(n^{-3/2}),
		\end{aligned}
		\]
		uniformly in $t$. Consequently,
		\[
		\mathrm{E}\left|
		\sum_{t=1}^{n}R_{nt}
		\right|
		\leq
		\sum_{t=1}^{n}E|R_{nt}|
		=
		O(n^{-1/2})
		\longrightarrow0.
		\]
		By Markov's inequality,
		$
		\sum_{t=1}^{n}R_{nt}
		=
		o_P(1),
		$
		which proves \eqref{srnt}, and hence \eqref{zn}.
		
		Since $\boldsymbol{\Lambda}_n\to\boldsymbol{\Lambda}$ with
		$\boldsymbol{\Lambda}$ positive definite and
		$\boldsymbol{\zeta}_n=O_P(1)$ under \textnormal{(A6)}, the minimizer
		$\boldsymbol{\Lambda}_n^{-1}\boldsymbol{\zeta}_n$ of
		$\widetilde Z_n$ is $O_P(1)$. By the convexity lemma and the argument
		in the final part of the proof of Theorem~1 in Pollard (1991), %\citet{pollard1991asymptotics}, 
		the pointwise approximation \eqref{zn} extends uniformly over compact subsets of
		$\mathbb{R}^{p}$, and
		$
		\hat{\boldsymbol{\delta}}_n
		-
		\boldsymbol{\Lambda}_n^{-1}
		\boldsymbol{\zeta}_n
		=
		o_P(1).
		$
		Therefore,
		\begin{equation}\label{ba}
			\sqrt{n}
			\left\{
			\hat{\boldsymbol{\beta}}_n(\omega_{\nu},\alpha,\psi)
			-
			\boldsymbol{\beta}_0(\alpha,\psi)
			\right\}
			=
			\boldsymbol{\Lambda}_n^{-1}(\alpha,\psi)
			\boldsymbol{\zeta}_n(\alpha,\psi)
			+
			o_P(1).
		\end{equation}
		
		For the general case $q>1$, apply the preceding argument separately
		to each of the $q$ regression problems. Since $q$ is fixed, stacking
		the resulting expansions gives
		\[
		\sqrt{n}\,
		\operatorname{vec}
		\left[
		\hat{\boldsymbol{\beta}}_{jn}(\omega_{\nu_j},\alpha,\psi)
		-
		\boldsymbol{\beta}_0(\alpha,\psi)
		\right]_{j=1}^{q}
		=
		\operatorname{vec}
		\left[
		\boldsymbol{\Lambda}_{jn}^{-1}(\alpha,\psi)
		\boldsymbol{\zeta}_{jn}(\alpha,\psi)
		\right]_{j=1}^{q}
		+
		o_P(1).
		\]
		By \textnormal{(A4)},
		$\boldsymbol{\Lambda}_{jn}(\alpha,\psi)
		\to\boldsymbol{\Lambda}_{j}(\alpha,\psi)$, while
		\textnormal{(A6)} gives the joint central limit theorem for
		$\operatorname{vec}[\boldsymbol{\zeta}_{jn}]_{j=1}^{q}$. Hence,
		Slutsky's theorem yields
		\[
		\sqrt{n}\,
		\operatorname{vec}
		\left[
		\hat{\boldsymbol{\beta}}_{jn}(\omega_{\nu_j},\alpha,\psi)
		-
		\boldsymbol{\beta}_0(\alpha,\psi)
		\right]_{j=1}^{q}
		\xrightarrow{D}
		N\left(
		\mathbf{0},
		\left[
		\boldsymbol{\Lambda}_{j}^{-1}(\alpha,\psi)
		\mathbf{V}_{jl}(\alpha,\psi)
		\boldsymbol{\Lambda}_{l}^{-1}(\alpha,\psi)
		\right]_{j,l=1}^{q}
		\right).
		\]
		This completes the proof of Lemma~1.

		\subsection*{A.2 Proof of Theorem 1}\label{pt1}
		
		For fixed $q\ge1$ and
		$0<\lambda_1<\cdots<\lambda_q<\pi$, let
		$\omega_{\nu_1},\ldots,\omega_{\nu_q}$ be Fourier frequencies
		satisfying
		\[
		\omega_{\nu_j}\longrightarrow\lambda_j,
		\qquad
		j=1,\ldots,q.
		\]
		For each $j$, define
		$
		\mathbf{x}_{jnt}
		=
		\left[
		1,
		\cos(\omega_{\nu_j}t),
		\sin(\omega_{\nu_j}t)
		\right]^{\top}.
		$
		The trigonometric regressors are uniformly bounded, so
		\textnormal{(A2)} holds. Moreover, Fourier orthogonality gives
		\[
		\mathbf{D}_{jn}
		=
		n^{-1}
		\sum_{t=1}^{n}
		\mathbf{x}_{jnt}\mathbf{x}_{jnt}^{\top}
		\longrightarrow
		\mathbf{D}_j
		=
		\operatorname{diag}
		\left\{
		1,\frac12,\frac12
		\right\},
		\]
		and hence \textnormal{(A4)} holds. Since
		$
		\boldsymbol{\Lambda}_{j}(\alpha,\psi)
		=
		\eta^{-1}(\alpha,\psi)\mathbf{D}_{j},
		$
		we have
		$
		\boldsymbol{\Lambda}_{j}^{-1}(\alpha,\psi)
		=
		\eta(\alpha,\psi)
		\operatorname{diag}\{1,2,2\}.
		$
		
		We next calculate the limiting covariance matrices. By
		\eqref{vnj}, when $j=l$,
		\[
		\mathbf{V}_{jjn}(\alpha,\psi)
		=
		\sum_{|\tau|<n}
		\gamma(\tau,\alpha,\psi)
		\left[
		n^{-1}
		\sum_{t=\max(1,1+\tau)}^{\min(n,n+\tau)}
		\mathbf{x}_{jnt}
		\mathbf{x}_{jn,t-\tau}^{\top}
		\right].
		\]
		For every fixed $\tau$,
		\[
		n^{-1}
		\sum_{t=\max(1,1+\tau)}^{\min(n,n+\tau)}
		\mathbf{x}_{jnt}
		\mathbf{x}_{jn,t-\tau}^{\top}
		\longrightarrow
		\operatorname{blockdiag}
		\left\{
		1,\frac12\mathbf{S}_{j}(\tau)
		\right\},
		\]
		where
		\[
		\mathbf{S}_{j}(\tau)
		=
		\begin{pmatrix}
			\cos(\lambda_j\tau)&-\sin(\lambda_j\tau)\\
			\sin(\lambda_j\tau)& \cos(\lambda_j\tau)
		\end{pmatrix}.
		\]
		Since $\gamma(\tau,\alpha,\psi)$ is an autocovariance function,
		it is symmetric in $\tau$. Therefore,
		\[
		\sum_{\tau=-\infty}^{\infty}
		\gamma(\tau,\alpha,\psi)
		\sin(\lambda_j\tau)
		=
		0.
		\]
		
		For simplicity, let
		$
		g_0:=g(0,\alpha,\psi)
		$
		and
		$
		g_j:=g(\lambda_j,\alpha,\psi).
		$
		Using the absolute summability of $\gamma$ and the definition of the
		AHS, the dominated convergence theorem gives
		\begin{equation}\label{valpha}
			\begin{aligned}
				\mathbf{V}_{jj}(\alpha,\psi)
				:=
				\lim_{n\to\infty}
				\mathbf{V}_{jjn}(\alpha,\psi)
				=
				\frac{1}{\eta^2(\alpha,\psi)}
				\operatorname{diag}
				\left\{
				g_0,
				\frac12 g_j,
				\frac12 g_j
				\right\}.
			\end{aligned}
		\end{equation}
		
		When $j\neq l$,
		\[
		\mathbf{V}_{jln}(\alpha,\psi)
		=
		\sum_{|\tau|<n}
		\gamma(\tau,\alpha,\psi)
		\left[
		n^{-1}
		\sum_{t=\max(1,1+\tau)}^{\min(n,n+\tau)}
		\mathbf{x}_{jnt}
		\mathbf{x}_{ln,t-\tau}^{\top}
		\right].
		\]
		Let
		$
		\mathbf{e}_1=(1,0,0)^{\top}.
		$
		Because the intercept components are common across frequencies,
		whereas the sine and cosine components at distinct limiting
		frequencies are orthogonal, for every fixed $\tau$,
		\[
		n^{-1}
		\sum_{t=\max(1,1+\tau)}^{\min(n,n+\tau)}
		\mathbf{x}_{jnt}
		\mathbf{x}_{ln,t-\tau}^{\top}
		\longrightarrow
		\mathbf{e}_1\mathbf{e}_1^{\top}.
		\]
		Consequently,
		\begin{equation}\label{valpha2}
			\begin{aligned}
				\mathbf{V}_{jl}(\alpha,\psi)
				:=
				\lim_{n\to\infty}
				\mathbf{V}_{jln}(\alpha,\psi)
				=
				\sum_{\tau=-\infty}^{\infty}
				\gamma(\tau,\alpha,\psi)
				\mathbf{e}_1\mathbf{e}_1^{\top}
				=
				\frac{g_0}
				{\eta^2(\alpha,\psi)}
				\mathbf{e}_1\mathbf{e}_1^{\top},
				\qquad j\neq l.
			\end{aligned}
		\end{equation}
		
		Combining \eqref{valpha} and \eqref{valpha2}, we may write
		\[
		\mathbf{V}_{jl}(\alpha,\psi)
		=
		\frac{1}{\eta^2(\alpha,\psi)}
		\left[
		g_0\mathbf{e}_1\mathbf{e}_1^{\top}
		+
		\mathbf{1}(j=l)
		\operatorname{diag}
		\left\{
		0,\frac12g_j,\frac12g_j
		\right\}
		\right].
		\]
		It follows that the $(j,l)$ block of the asymptotic covariance matrix
		of the AHR estimators is
		\[
		\begin{aligned}
			\boldsymbol{\Sigma}_{jl}
			&:=
			\boldsymbol{\Lambda}_{j}^{-1}(\alpha,\psi)
			\mathbf{V}_{jl}(\alpha,\psi)
			\boldsymbol{\Lambda}_{l}^{-1}(\alpha,\psi)\\
			&=
			g_0\mathbf{e}_1\mathbf{e}_1^{\top}
			+
			\mathbf{1}(j=l)
			\operatorname{diag}
			\left\{
			0,2g_j,2g_j
			\right\}.
		\end{aligned}
		\]
		Thus, the full coefficient vectors at distinct frequencies share the
		same asymptotic intercept component, but their harmonic components
		are asymptotically uncorrelated.
		
		Condition \textnormal{(C1)} verifies \textnormal{(A1)}, the boundedness
		of the trigonometric regressors verifies \textnormal{(A2)}, and
		\textnormal{(C2)} implies \textnormal{(A3)}. The convergence of the
		design matrices verifies \textnormal{(A4)}, the covariance
		calculations above verify \textnormal{(A5)}, and
		\textnormal{(A6)} follows directly from \textnormal{(C4)}. Therefore,
		Lemma~1 gives
		\[
		\sqrt{n}\,
		\operatorname{vec}
		\left[
		\hat{\boldsymbol{\beta}}_{jn}
		(\omega_{\nu_j},\alpha,\psi)
		-
		\boldsymbol{\beta}_0(\alpha,\psi)
		\right]_{j=1}^{q}
		\xrightarrow{D}
		N\left(
		\mathbf{0},
		[\boldsymbol{\Sigma}_{jl}]_{j,l=1}^{q}
		\right).
		\]
		
		Let
		\[
		\mathbf{R}
		=
		\begin{pmatrix}
			0&1&0\\
			0&0&1
		\end{pmatrix},
		\]
		which selects the cosine and sine coefficients. Since
		$
		\mathbf{R}\mathbf{e}_1=\mathbf{0},
		$
		we have
		$
		\mathbf{R}
		\boldsymbol{\Sigma}_{jl}
		\mathbf{R}^{\top}
		=
		2g_j\mathbf{I}_2\mathbf{1}(j=l).
		$
		Consequently,
		\[
		\sqrt{n}\,
		\operatorname{vec}
		\left[
		\mathbf{R}
		\left\{
		\hat{\boldsymbol{\beta}}_{jn}
		(\omega_{\nu_j},\alpha,\psi)
		-
		\boldsymbol{\beta}_0(\alpha,\psi)
		\right\}
		\right]_{j=1}^{q}
		\xrightarrow{D}
		N\left(
		\mathbf{0},
		\operatorname{blockdiag}
		\left\{
		2g_1\mathbf{I}_2,\ldots,
		2g_q\mathbf{I}_2
		\right\}
		\right).
		\]
		Because
		$
		\mathbf{R}\boldsymbol{\beta}_0(\alpha,\psi)=\mathbf{0},
		$
		the vectors
		\[
		\frac{
			\sqrt{n}\,
			\mathbf{R}
			\hat{\boldsymbol{\beta}}_{jn}
			(\omega_{\nu_j},\alpha,\psi)
		}{
			\sqrt{2g_j}
		},
		\qquad j=1,\ldots,q,
		\]
		converge jointly to independent
		$N(\mathbf{0},\mathbf{I}_2)$ random vectors.
		
		Define
		$
		\boldsymbol{\xi}_j
		=
		(\xi_{1j},\xi_{2j})^{\top},
		\quad
		j=1,\ldots,q,
		$
		where the components are independent standard normal random variables.
		By the continuous mapping theorem,
		\[
		\begin{aligned}
			\frac{
				\operatorname{AHP}_n
				(\omega_{\nu_j},\alpha,\psi)
			}{
				g_j
			}
			&=
			\frac{n}{
				4g_j
			}
			\left\{
			\hat\beta_{2,jn}^{\,2}
			(\omega_{\nu_j},\alpha,\psi)
			+
			\hat\beta_{3,jn}^{\,2}
			(\omega_{\nu_j},\alpha,\psi)
			\right\}
			&\xrightarrow{D}
			\frac12
			\left(
			\xi_{1j}^{2}+\xi_{2j}^{2}
			\right)
			=
			\frac12\chi_{2,j}^{2}.
		\end{aligned}
		\]
		Since the limiting harmonic coefficient vectors are independent across
		$j$, the variables
		$\chi_{2,1}^{2},\ldots,\chi_{2,q}^{2}$ are independent. Therefore,
		\[
		\left\{
		\operatorname{AHP}_n
		(\omega_{\nu_1},\alpha,\psi),
		\ldots,
		\operatorname{AHP}_n
		(\omega_{\nu_q},\alpha,\psi)
		\right\}
		\xrightarrow{D}
		\left\{
		\frac12
		g_1\chi_{2,1}^{2},
		\ldots,
		\frac12
		g_q\chi_{2,q}^{2}
		\right\},
		\]
		which proves \eqref{the2}.
	\end{proof}
	
	\section*{A.3 Proof of Theorem 2}
	%	First, by (C5)--(C6), averaging across $M_n$ nearby Fourier frequencies reduces the variance by a factor of order $1/(M_n)$ while introducing only $o(1)$ bias because the true expectile spectrum $g(\lambda,\alpha)$ is continuous in $\lambda$ under assumption (C4).  Second, a central limit theorem for quadratic forms of mixing processes \citep{brillinger2001time, dahlhaus1988empirical, davidson1994stochastic} implies that
	%\begin{equation}\label{q4}
	%	\sqrt{M_n}\big(	\hat{g}(\omega_{\nu}, \alpha,\psi)  -  g(\lambda, \alpha,\psi)\big) \overset{d}{\longrightarrow} N(0, \|K^*\|_2^2 2g(\lambda,\alpha,\psi)).
	%\end{equation}
	%Since both bias and variance vanish under (C4)--(C6), we obtain the consistency in (\ref{p3}) and the asymptotic distribution in \eqref{q4}.
	
	\begin{proof}
		For notational simplicity, write
		\[
		A_{n,s}
		:=
		\operatorname{AHP}_n
		(\omega_{\nu+s},\alpha,\psi),
		\qquad
		g_{n,s}
		:=
		g(\omega_{\nu+s},\alpha,\psi).
		\]
		Then
		\[
		\hat g(\omega_\nu,\alpha,\psi)
		=
		\sum_{s=-M_n}^{M_n}W_{n,s}A_{n,s}.
		\]
		
		We first examine the frequencies within the smoothing window.
		By condition \textnormal{(C6)},
		$
		\omega_\nu\rightarrow\lambda,
		\frac{M_n}{n}\rightarrow0.
		$
		Since
		$
		\omega_{\nu+s}-\omega_\nu
		=
		\frac{2\pi s}{n},
		$
		we have
		\[
		\begin{aligned}
			\max_{|s|\leq M_n}
			|\omega_{\nu+s}-\lambda|
			&\leq
			|\omega_\nu-\lambda|
			+
			\max_{|s|\leq M_n}
			|\omega_{\nu+s}-\omega_\nu|\\
			&\leq
			|\omega_\nu-\lambda|
			+
			\frac{2\pi M_n}{n}
			\longrightarrow0.
		\end{aligned}
		\]
		Thus, all frequencies in the smoothing window converge uniformly to
		the interior frequency $\lambda$.
		
		By condition \textnormal{(C5)}, the function
		$g(\cdot,\alpha,\psi)$ is continuous at $\lambda$. Consequently,
		\begin{equation}\label{uniform-g}
			\sup_{|s|\leq M_n}
			\left|
			g_{n,s}-g(\lambda,\alpha,\psi)
			\right|
			\longrightarrow0.
		\end{equation}
		
		Conditions \textnormal{(C1)--(C4)} are used through Theorem~1.
		Specifically, they imply the limiting chi-square distribution of the
		raw AHP along every sequence of interior Fourier frequencies
		converging to $\lambda$. The additional moment condition \eqref{moment_condition}
		ensures uniform integrability of the corresponding squared AHP
		ordinates. Therefore, convergence in distribution can be upgraded to
		convergence of the first two moments.
		
		We next show that these moment convergences hold uniformly over the
		smoothing window. In particular,
		\begin{equation}\label{uniform-mean}
			\sup_{|s|\leq M_n}
			\left|
			E(A_{n,s})-g_{n,s}
			\right|
			\longrightarrow0
		\end{equation}
		and
		\begin{equation}\label{uniform-var}
			\sup_{|s|\leq M_n}
			\left|
			\operatorname{Var}(A_{n,s})-g_{n,s}^{2}
			\right|
			\longrightarrow0.
		\end{equation}
		
		To verify \eqref{uniform-mean}, suppose, to the contrary, that it does
		not hold. Then there exist $\epsilon_0>0$, a subsequence
		$\{n_k\}$, and indices $s_k$ satisfying $|s_k|\leq M_{n_k}$ such
		that
		$
		\left|
		E(A_{n_k,s_k})-g_{n_k,s_k}
		\right|
		\geq\epsilon_0.
		$
		Condition \textnormal{(C6)} implies
		$
		\omega_{\nu+s_k}\longrightarrow\lambda.
		$
		By Theorem~1 and the moment condition \eqref{moment_condition},
		$
		E(A_{n_k,s_k})
		\longrightarrow
		g(\lambda,\alpha,\psi).
		$
		On the other hand, \eqref{uniform-g} gives
		$
		g_{n_k,s_k}
		\longrightarrow
		g(\lambda,\alpha,\psi),
		$
		which is a contradiction. Hence \eqref{uniform-mean} holds. The same
		subsequence argument, using convergence of the second moments, proves
		\eqref{uniform-var}.
		
		We now consider the bias. By the nonnegativity and normalization of
		the weights in condition \textnormal{(C7)},
		$
		W_{n,s}\geq0,
		\sum_{s=-M_n}^{M_n}W_{n,s}=1.
		$
		Therefore,
		\[
		\begin{aligned}
			\left|
			\mathrm{E}\left\{
			\hat g(\omega_\nu,\alpha,\psi)
			\right\}
			-
			g(\lambda,\alpha,\psi)
			\right|
			\leq{}&
			\sum_{s=-M_n}^{M_n}
			W_{n,s}
			\left|
			E(A_{n,s})-g_{n,s}
			\right|
			+
			\sum_{s=-M_n}^{M_n}
			W_{n,s}
			\left|
			g_{n,s}-g(\lambda,\alpha,\psi)
			\right|\\
			\leq{}&
			\sup_{|s|\leq M_n}
			\left|
			\mathrm{E}(A_{n,s})-g_{n,s}
			\right|
			+
			\sup_{|s|\leq M_n}
			\left|
			g_{n,s}-g(\lambda,\alpha,\psi)
			\right|.
		\end{aligned}
		\]
		It follows from \eqref{uniform-g} and \eqref{uniform-mean} that
		\begin{equation}\label{bias-zero}
			\mathrm{E}\left\{
			\hat g(\omega_\nu,\alpha,\psi)
			\right\}
			-
			g(\lambda,\alpha,\psi)
			\longrightarrow0.
		\end{equation}

		Next, decompose the variance as
		\[
		\begin{aligned}
			\operatorname{Var}
			\left\{
			\hat g(\omega_\nu,\alpha,\psi)
			\right\}
			={}
			\sum_{s=-M_n}^{M_n}
			W_{n,s}^{2}\operatorname{Var}(A_{n,s}) 
			+
			\sum_{\substack{|s|\leq M_n,\ |r|\leq M_n\\ s\neq r}}
			W_{n,s}W_{n,r}
			\operatorname{Cov}(A_{n,s},A_{n,r}).
		\end{aligned}
		\]
		
		By \eqref{uniform-g} and \eqref{uniform-var},
		$
		\sup_{|s|\leq M_n}
		\operatorname{Var}(A_{n,s})
		=
		g^{2}(\lambda,\alpha,\psi)+o(1).
		$
		Consequently,
		\[
		\begin{aligned}
			\sum_{s=-M_n}^{M_n}
			W_{n,s}^{2}\operatorname{Var}(A_{n,s})
			\leq
			\left\{
			g^{2}(\lambda,\alpha,\psi)+o(1)
			\right\}
			\sum_{s=-M_n}^{M_n}W_{n,s}^{2} 
			\leq
			\left\{
			g^{2}(\lambda,\alpha,\psi)+o(1)
			\right\}
			\max_{|s|\leq M_n}W_{n,s},
		\end{aligned}
		\]
		where the second inequality follows from the nonnegativity and
		normalization of the weights:
		\[
		\sum_{s=-M_n}^{M_n}W_{n,s}^{2}
		\leq
		\left(
		\max_{|s|\leq M_n}W_{n,s}
		\right)
		\sum_{s=-M_n}^{M_n}W_{n,s}
		=
		\max_{|s|\leq M_n}W_{n,s}.
		\]
		Since
		$
		\max_{|s|\leq M_n}W_{n,s}
		=
		O(M_n^{-1})
		\longrightarrow 0
		$
		by conditions \textnormal{(C6)} and \textnormal{(C7)}, it follows that
		\begin{equation}\label{diagonal-zero}
			\sum_{s=-M_n}^{M_n}
			W_{n,s}^{2}\operatorname{Var}(A_{n,s})
			\longrightarrow 0.
		\end{equation}

		For the off-diagonal contribution, condition \textnormal{(C8)}
		implies
		\[
		\begin{aligned}
			\left|
			\sum_{\substack{|s|\leq M_n,\ |r|\leq M_n\\s\neq r}}
			W_{n,s}W_{n,r}
			\operatorname{Cov}(A_{n,s},A_{n,r})
			\right|
			\qquad\leq
			\sum_{\substack{|s|\leq M_n,\ |r|\leq M_n\\s\neq r}}
			W_{n,s}W_{n,r}
			\left|
			\operatorname{Cov}(A_{n,s},A_{n,r})
			\right|
			\longrightarrow0.
		\end{aligned}
		\]
		Combining this result with \eqref{diagonal-zero} yields
		$
		\operatorname{Var}
		\left\{
		\hat g(\omega_\nu,\alpha,\psi)
		\right\}
		\rightarrow0.
		$
		Then, by \eqref{bias-zero} and Chebyshev's inequality,
		$
		\hat g(\omega_\nu,\alpha,\psi)
		-
		g(\lambda,\alpha,\psi)
		\xrightarrow{p}0.
		$
		Therefore,
		$
		\hat g(\omega_\nu,\alpha,\psi)
		\xrightarrow{p}
		g(\lambda,\alpha,\psi).
		$
	\end{proof}
\end{document}